\documentclass{aastex}
\usepackage{spr-astr-addons}
\usepackage{url}
\RequirePackage{color}

\pdfoutput=1

\begin{document}

\title{Rapid Mass segregation in small stellar clusters}
\shorttitle{Rapid mass segregation}
\shortauthors{Spera & Capuzzo-Dolcetta}

\author{Mario Spera\altaffilmark{1,3}} \and \author{Roberto Capuzzo-Dolcetta \altaffilmark{2}}
\altaffiltext{1}{Dep. of Physics, Universit\'a degli Studi di Milano Bicocca, Italy}
\altaffiltext{2}{Dep. of Physics, Sapienza, Universit\'a di Roma, Italy}
\altaffiltext{3}{Institut f\" ur Astro- und Teilchenphysik – Universit\" at Innsbruck, Austria}

\begin{abstract}
In this paper we focus our attention on small-to-intermediate $N$-body systems that are, initially, distributed uniformly in space and dynamically~\lq cool\rq~(virial ratios $Q=2T/|\Omega|$ below $\sim 0.3$). In this work, we study the mass segregation that emerges after the initial violent dynamical evolution. At this scope, we ran a set of high precision $N$-body simulations of isolated clusters by means of \texttt{HiGPUs}, our direct summation $N$-body code. After the collapse, the system shows a clear mass segregation. This (quick) mass segregation occurs in two phases: the first shows up in clumps originated by sub-fragmentation before the deep overall collapse; this segregation is partly erased during the deep collapse to re-emerge, abruptly, during the second phase, that follows the first bounce of the system. In this second stage, the proper clock to measure the rate of segregation is the dynamical time after virialization, which (for cold and cool systems) may be significantly different from the crossing time evaluated from initial conditions. This result is obtained for isolated clusters composed of stars of two different masses (in the ratio $m_h/m_l=2$), at varying their number ratio, and is confirmed also in presence of a massive central object (simulating a black hole of stellar size). Actually, in stellar systems starting their dynamical evolution from cool conditions, the fast mass segregation adds to the following, slow, secular segregation which is collisionally induced. The violent mass segregation is an effect persistent over the whole range of $N$ ($128 \leq N \leq 1024$) investigated, and is an interesting feature on the astronomical-observational side, too. 
The semi-steady state reached after virialization corresponds to a mass segregated distribution function rather than that of  equipartition of kinetic energy per unit mass as  it should result from violent relaxation.
\end{abstract}

\keywords{Methods: numerical \and Galaxies: star clusters: general \and Galaxies: kinematics and dynamics}


\section{Introduction}
\label{intro}
Astronomical observations indicate that several stellar systems (from young and very young open star clusters to rich clusters of galaxies) show a certain degree of segregation of the most luminous and massive components in their inner regions \citep{bontemps2010,kirk2014,er2013,gouliermis2006,raboud1998,raboud1999,littlefair2003}. For instance, the Orion Nebula Cluster (ONC) is mass segregated down to about $5$ $\mathrm{M}_{\odot}$ despite its young age, estimated to be less than 2 Myr \citep{hillen1998,allison2009}. 
In recent times, many papers have been dedicated to the study of both short and long term evolution of star clusters of different sizes both in isolation and in presence of a tidal field or a residual gas 
\citep{ban15,far15,ale16,ros17,sha17,yu17}. 
The dominant cause of the rapid mass segregation process, for such systems, is still under debate. 
A dynamical origin is commonly excluded; in fact, many of the systems which show a mass segregation are much younger than their two-body relaxation time, which is usually considered as the time needed to segregate masses. 

As well known, the relaxation time can be written as

\begin{equation}
\label{eq:relaxation}
t_{rel} = \frac{v^2\sigma X}{4\sqrt{2}\pi G^2 \overline{\rho} ~\overline{m} \ln \Lambda F(X)}
\end{equation}

where $v$ is the typical velocity of a star in the system, $\sigma$ the system velocity dispersion, $X\equiv v/(\sqrt{2}\sigma)$, $G$ the gravitational constant, $\overline{\rho}$  the mean mass density of the field (target) stars, $\overline{m}$ their mean mass, $\ln \Lambda$ the usual Coulomb logarithm, and $F(X)$ is the function

\begin{equation}
\label{eq:gxfokker}
F(X)=\frac{1}{2X^2}\left[\mathrm{erf}(X)-\frac{2}{\sqrt{\pi}}Xe^{-X^2}\right].
\end{equation}

Equation (\ref{eq:relaxation}) is valid whether the initial conditions are, indeed, not too far from virial equilibrium (see for example \citet{binney2008}). This is not always a correct assumption, especially if we refer to the study of the early dynamical evolution of stellar systems whose stars form in sub-structured, clumpy regions in sub-virial conditions \citep{demarchi2013,schmeja2006,adams2006}.
It has been already shown \citep{farouki1982,allison2009,mcmillan2007}, through $N$-body simulations, that, if the initial state of a stellar system is out of equilibrium and the initial spatial distribution of its stars is not homogeneous, a significant degree of mass segregation can emerge on very short time-scales, significantly shorter than the two-body relaxation time.
\citet{allison2009} argue that the main mechanism that leads to mass segregation on a short time is related to the short interval of time in which the violent collapse creates a very dense core; typically, it contains about half the mass of the stellar system in a radius of about one tenth of its initial size. They show that the time to segregate masses down to $4 - 5$ $\mathrm{M}_{\odot}$, in the dense core, is comparable to its living time (approximatively $0.1$ Myr). This is enough to justify also the degree of mass segregation observed in some astrophysical systems, such as the cited ONC (\citet{allison2009b}).
On the other hand, \citet{bonnell1998} excluded that the observed mass segregation in young stellar clusters could be due to a violent dynamical evolution, introducing, rather, the hypothesis of an in situ formation of the most massive and luminous stars. In their work, they investigated the dynamical evolution of both spherical stellar systems, initially in virial equilibrium, and non-spherical stellar systems in sub-virial conditions. They found that the time-scale for mass segregation was largely unaffected by differences in the initial phase-space distribution of the stars.

A paper by \citet{cap14} investigated the role of initial sub-virial conditions on the final state of a set of stellar system on a range of $N$ from $2,048$ to $131,072$ and intermediate mass black hole therein and found, by a match of results with some observational data, that the initial values of the virial ratio $Q=2T/|\Omega|$
(where $T$ and $\Omega$ are, respectively, the system's total kinetic and potential energies) should be in the range $0.36$-$0.50$, excluding cold initial conditions. They also found an enhanced mass segregation in initially cooler systems but they give no evidence of sub-clustering before the collapse. 
It is noteworthy that the clusters studied in that last study never reach equilibrium over the time-span explored.

In the light of the described scenario, it is clear that a precise interpretation of rapid mass segregation has not been reached, yet. It is actually difficult to discriminate between the two above formulated origins of the quick segregation of masses in young star clusters. 

In the attempt of intepreting the mass segregation observed in young stellar clusters it must be kept in mind that the evolutionary ~\lq clock\rq~ which properly measures the rate of internal evolution of the cluster after virialization may change significantly when dealing with systems initially cold (i.e. with small and very small virial ratios). Actually, in these cases, the initial crossing time does not mean much because the system virializes in a short time (few initial crossing times) to a dense state which naturally speeds up the internal cluster evolution.

Moreover, an additional problem comes from the 
observational side, because very young star clusters, still sites of star formation processes, are very 
arduous to observe because they are embedded in high density gas clouds. Therefore, it is hard to prove if 
the most massive stars form, indeed, very close to the innermost regions or, on the contrary, they form on a 
larger spatial scale and segregate later.

In this context, a dynamical mechanism, not deeply investigated yet (even if already highlighted by \citet{aarseth1988}, \citet{mcmillan2015}) and that might play an important role in segregating masses, is the initial rapid fragmentation of a stellar systems whose stars are initially distributed homogeneously with very small initial velocities. \citet{mcmillan2007} and \citet{mcmillan2015} stressed that several sub-systems form during the collapse phase; in particular, they show, just before the bounce of the system, a certain degree of mass segregation which is preserved after the sequent collapse and rebound. It is important to stress that  this kind of segregation occurs before the formation of the short-living core described in \citet{allison2009}.
 
Given all this, we are forced to conclude that the theoretical scenario is still unclear and deserves investigation. Moreover, the real situation is made even more complicated by the presence of a certain number of primordial binaries and/or of black holes, and/or of a background gas. The effect of an external gas has been actually considered in this paper but just in a simplified scheme of an external, analytic time varying potential, starting from pioneering work by \citet{hil80} followed by simulations of $N-$ body evolutions in an an expanding potential in a more or less complicated environment \citet{lad84,gey01,ban15}. So far, no fully, self-consistent $N-$ body + hydrodynamics simulations have been done to examine this phase. 

To provide some advancement in the topic described above,  in this paper we study the effects of the violent collapse of an $N$-body self-gravitating system starting from initial conditions corresponding to a virial ratio $0\leq Q \leq 1$ with stars uniformly distributed in space. 
The precise aim is to identify the times and modes of mass segregation in a cluster of moderate size in a clear set of initial conditions and a clear lay out of the approximations assumed. We also check the role played by the presence of both a residual gas after star formation (although in a simple external gas approximation) and the presence of a black hole of stellar origin. When dealing with systems containing a massive object (likely, a black hole as remnant of a progenitor massive enough to end its life within few Myr), we assume it was present since the start of the evolution, which is only approximately correct because, although the black hole mass we adopt is high (see Sect. \ref{sec2.1}), its progenitor star evolutionary lifetime is not zero but of the order of 3-4 Myr.  

In all our $N$-body simulations, where we privileged the clarity of results respect to their generality, we 
consider the simple case of a bimodal mass spectrum with bodies initially distributed randomly in a sphere of 
unitary initial radius $R$. 

The organization of the paper is: in Sect. \ref{sec:model} we describe the models of stellar systems we adopted and give a description of both the software and hardware resources used to follow their dynamical evolution; in Sect. \ref{sec:results} we present and discuss the results with attention to both the physics of violently relaxing, intermediate $N$, systems and to the possible comparison with observational data of real clusters. In Sect. \ref{sec:disc} we discuss the results. Finally, in Sect. \ref{sec:concl} we list our conclusions and outline the developments needed to get a better insight into the evolution of open clusters emerging from their mother proto-clouds.

\section{Model}
\label{sec:model}
\subsection{Stellar systems}
\label{sec2.1}
We performed a large set of direct $N$-body simulations of young star clusters with $128\leq N\leq1,024$. 

The pair interaction is via a softened Newtonian potential, 
$\Phi_{ij} \propto (r_{ij}^2+\epsilon^2)^{-1/2}$, with a softening length  assumed $\epsilon =10^{-5}R$ where $R$ is the initial radial extension of the system. This value preserves  the Newtonian behaviour being significantly smaller than the average closest-neighbour distance $\sim N^{-1/3}$.
Actually, we checked that the softening length assumed is sufficiently small to allow the, transient, formation of binaries during the high compression phase reached by ~\lq cool\rq~ systems.  

We adopted a bimodal mass spectrum: ~\lq light\rq~
\par\noindent
stars (each with mass $m_l$ and total number $N_l$) and ~\lq heavy\rq~   stars (each with mass $m_h$ and total number $N_h$) such that $m_h = 2m_l$ and $N_h=N_l$. The initial spatial coordinates of bodies  have been sampled from a uniform distribution in a sphere of radius $R=1$. The initial velocities have random orientations and, in order to determine their absolute values, we assume a certain value of the initial virial ratio $Q$, already defined as

\begin{equation}
Q=\frac{2T}{\left| \Omega \right|}
\end{equation}

remembering that $T$ is the kinetic energy and $\Omega$ the potential energy of the $N$-body system. We vary $Q$ from $Q=0$, corresponding to the most violent (cold) collapse, up to $Q=1$, which means virial equilibrium, at steps of $0.1$. Small values of $Q$ (\lq cool\rq~ systems)  seem to be suggested by observations of young stellar systems where stars appear to be clustered in sub-virial conditions (see, for example, \citet{adams2006} and \citet{hillen1998}).
We also investigated the gravitational role of  a residual gas, modelled as an analytical contribution to the accelerations of stars. This field  is represented, at any time $t$, by a time varying Plummer potential \citep{plummer1911}
\begin{equation}
\Phi_P\left(r;t\right)=-\frac{GM_G}{\sqrt{r^2+a^2\left(t\right)}}
\label{eq:Plu}
\end{equation} 

where $M_G$ is the total (constant) gas mass and $a(t)$ is the Plummer model scale length (assumed time-dependent), which is connected to the core radius (defined as the distance to the center where the surface density halves its central value) by the relation $r_c= a\sqrt{\sqrt{2}-1}\simeq 0.644a$.
Another ingredient of our simulations is the inclusion of a particle with a mass significantly larger than the others. This object may be considered as a stellar black hole, whose mass is indicated as $m_{BH}$. The black hole mass was assumed to be  $m_{BH}=25m_h=50m_l$. The total mass of the star cluster system, $M_C$, is assumed as unit of mass, that is
\begin{equation}
M_C=N_h m_h + N_l m_l + M_G + m_{BH} = M_* + M_G = 1
\end{equation}
where $M_*$ refers to the total mass in stars. The mass of the gas the stellar system is embedded in is given by assuming a value for the star formation efficiency parameter, that is
\begin{equation}
\mathcal S=\frac{M_*}{M_C}=\frac{M_*}{M_G+M_*}.
\label{eq:sfe}
\end{equation}
Once given $\mathcal S$, the stellar mass of the cluster is $M_*=\mathcal SM_C$ and  $M_G=M_C(1-\mathcal S)$. We considered two values of $\mathcal S$,  $\mathcal S=0.3$ and $\mathcal S=1$ (no gas). Because we are interested in the emerging state of young open clusters, we evolved our stellar systems up to a relatively short time ($<5 \mathrm{Myr}$), therefore we neglected, in a first approximation, the effects of stellar evolution. For each set of values of the free parameters we performed 30 runs ($\gtrsim 5,000$ simulations, in total) to give a statistical significance to our results. 

Unless otherwise specified, the results of our simulations will be presented using the  initial crossing time of the system as time unit, given for the gravitational constant the value $G=1$. The velocity units are obtained consequently.

Table \ref{tab:tab1} summarizes the properties of our simulations.

\begin{table*}
	\caption{Summary of the simulations performed for this paper. Column 1:name of the simulation set. Column 2: range of number of stars. Column 3: ratio between the number of heavy and light stars. Column 4: presence or absence of a stellar-mass black hole. Column 5: presence or absence of a background gravitational potential. Column 6: number of simulations performed.}
	\label{tab:tab1}
	\begin{tabular}{cccccc}
		\hline
		Name & $N$ & $N_h/N_l$ & BH & Back. pot. & simulations n.\\
		\hline
		Case 1 & $\left[128-1,024\right]$ & 1 & NO & NO & 30 (for each $Q$)\\
		Case 2 & $\left[128-1,024\right]$ & 1 & YES & NO & 30 (for each $Q$)\\
		Case 3 & $\left[128-1,024\right]$ & 1 & NO & YES (static) & 30 (for each $Q$)\\
		Case 4 & $\left[128-1,024\right]$ & 1 & NO & YES (expanding) & 30 (for each $Q$)\\
		Case 5 & $1,024$ & $\left[0.125;1\right]$ & NO & NO & 30 ($Q=0$ only)\\
		\hline
	\end{tabular}
\end{table*}

\subsection{The $N$-body code}
For the purposes of this paper, we used our highly parallel $N$-body code \texttt{HiGPUs} \citep{dolcetta2013} running on our private machine containing a Central Processing Unit (CPU) Intel i7 950 and 2 nVIDIA Tesla C2050 (Fermi) Graphics Processing Units (GPUs). Although the GPUs used are not the best in terms of cost and computing capability, being old generation cards, thanks to the high single core working frequency they perform well in regime of weak load, i.e. using a number of particles $\lesssim 1,024$ \citep{dolcetta2013b}.
\texttt{HiGPUs} is based on a Hermite 6th order integration scheme \citep{nitadori2008} parallelized using CUDA (or OpenCL) plus OpenMP and MPI to guarantee maximum performance when a hybrid (CPUs+GPUs) computer is used. The integration algorithm is implemented using the technique of block time steps \citet{aarseth2003}. After proper speed/precision testing we decided to determine the particle time steps using the following formula:
\begin{equation}
\label{eq:criterion}
\Delta t_i=\frac{1}{\alpha_1+\alpha_2}\left[\alpha_1\eta_4\left(\frac{A_i^{(1)}}{A_i^{(2)}}\right) + \alpha_2\eta_6\left(\frac{A_i^{(1)}}{A_i^{(4)}}\right)^{\frac{1}{3}}\right],
\end{equation}where
\begin{equation}
A_i^{(s)} \equiv \sqrt{\left|\mathbf{a}_i^{(s-1)}\right|\left|\mathbf{a}_i^{(s+1)}\right|+\left|\mathbf{a}_i^{(s)}\right|^2}
\end{equation}
and $\mathbf{a}_i^{(s)}$ is the \textit{s}-th time derivative of the acceleration of the $i$-th particle. Equation \ref{eq:criterion} represents a weighted mean (with coefficients $\alpha_1$ and $\alpha_2$) between the Aarseth criterion for the 4th order Hermite integrator (with accuracy parameter $\eta_4$) \citep{aarseth2003} and the generalized Aarseth criterion for the 6th order scheme (with accuracy parameter $\eta_6$) \citep{nitadori2008}. The combination with $\alpha_1=\alpha_2=\frac{1}{2}$ has been found to be more stable, for the 6th order method, than the two criteria used independently, providing a better total energy conservation and avoiding time steps either too large or too small.

\section{Results}
\label{sec:results}
In the following we present and discuss mainly the results  for $N=1,024$, chosen as reference case. This is motivated by that similar results, in terms of mass segregation, are obtained in the whole range of $N$ here studied, $128\leq N \leq 1,024$, as shown above.

One good indicator to quantify the level of mass segregation is the ratio of the lagrangian radii of heavy particles to those of light particles, so that values significantly smaller than 1 indicate the presence of mass segregation. Of course, the use of lagrangian radii ratios as mass segregation indicators is reliable when the system has a spherical symmetry. Nevertheless, before the first bounce of the system, we observe a fragmentation of the system into various clumps which are not spherical. In such case, a more reliable evaluation of the presence of mass segregation comes from the so called {\em minimum spanning tree} method (MST)\footnote{The minimum spanning tree is the shortest path length which connects a certain number of points without forming closed loops.}, developed by \citet{allison2009}. Given a sub-set $K_m$ of $m$ points belonging to a system composed of $m'>m$ bodies, the degree of mass segregation established for that sample, $\Lambda_{K_m}$, is defined as 
\begin{equation}
\label{eq:MST}
\Lambda_{K_m} = \frac{\left<l_{norm}\right>}{l_{K_m}}\pm \frac{\sigma_{norm}}{l_{K_m}}
\end{equation}
where $l_{K_m}$ is the MST for the sample $K_m$, $\left<l_{norm}\right>$ is the average MST for $m$ randomly selected stars in the whole system and $\sigma_{norm}$ is its associated standard deviation. In each run, we calculate $\left<l_{norm}\right>$ using 200 different sub-sets of points. From the definition given in Eq. \ref{eq:MST} it follows that the generic sample $K$ is mass segregated if $\Lambda_{K_m}$ is significantly greater than 1. 

When we use the MST method to investigate the distribution of masses in our simulations, we need to remove escapers. This removal process is important because, after the first bounce, a significant amount of mass (about 20\% of the total mass) is lost; therefore, a single, remote, star may alter significantly the length of the spanning tree of a specific population.  To identify escapers correctly, the best criteria are those based upon both mechanical energy (which should be positive) and distance from the densest region of the system (the core). Consequently, we consider a star as lost from the cluster when it has positive energy and a distance from the cluster density center larger than twice the initial core radius, thus avoiding the possibility for the star to reduce again energy to negative values.

The simplest case we studied concerns systems with neither gas nor a central black hole. To give statistical 
significance to our results, we generated a set of 30 different initial configurations to evolve, by simply 
changing the seed of the Mersenne Twister random number generator \citep{matsumoto1998} that provides 
initial conditions in the phase space. Here we show the results obtained from these simulations.

\begin{figure}
\centering
\includegraphics[width=\columnwidth]{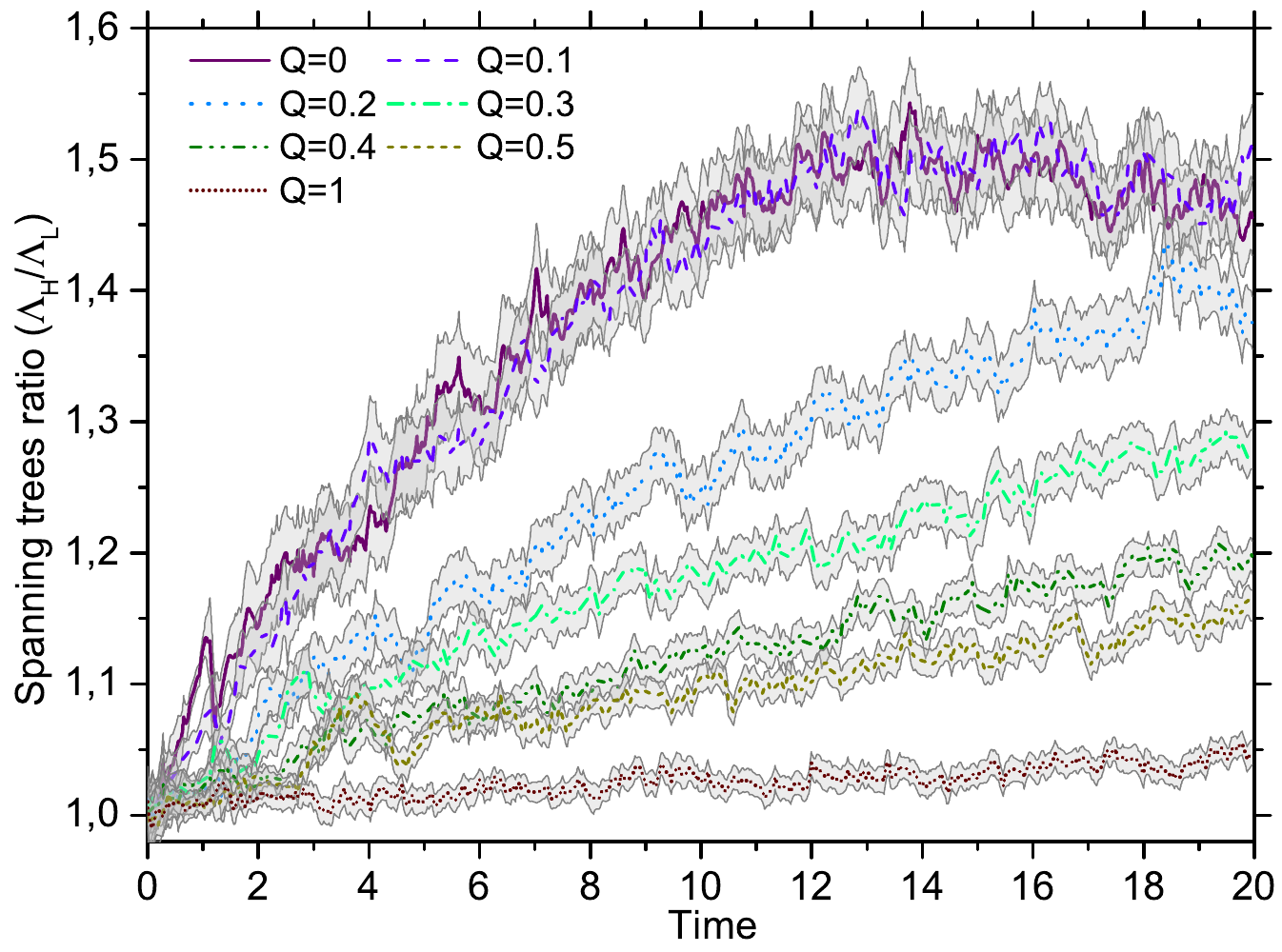}
\caption{The ratio between $\Lambda_h$ and $\Lambda_l$ for several values of the initial virial ratio $Q$, for the simulations of Case 1 with $N=1,024$ (see Tab. \ref{tab:tab1}). 
Values of $Q$ between $0.5$ and $0.9$ have also been considered, but they are not shown here for the sake of 
a clear representation of the results. Each curve represents an average value of $\Lambda_h / \Lambda_l$ and 
the error (standard deviation) is represented by the semi-transparent area.}
\label{fig:Fig1}
\end{figure}
Figure \ref{fig:Fig1} shows the ratio between the averaged value of $\Lambda$ for the heavy ($\Lambda_h$) and 
for the light ($\Lambda_l$) stars as a function of time. The first important thing to stress is that the 
degree of mass segregation depends strictly upon the initial state of the system: the farther from 
equilibrium, the higher and the quicker the degree of the resulting mass segregation. 

Another important result is that we obtain a significant degree of mass segregation starting 
from homogeneous and smooth initial conditions. For the cases of most violent collapses ($Q=0$ and $Q=0.1$) 
the system gets to a saturation of the degree of mass segregation around 12 units of time, while, for larger 
values of the initial virial ratio, the process of mass segregation continues at an approximatively constant 
rate.

\begin{figure}
\centering
\includegraphics[width=\columnwidth]{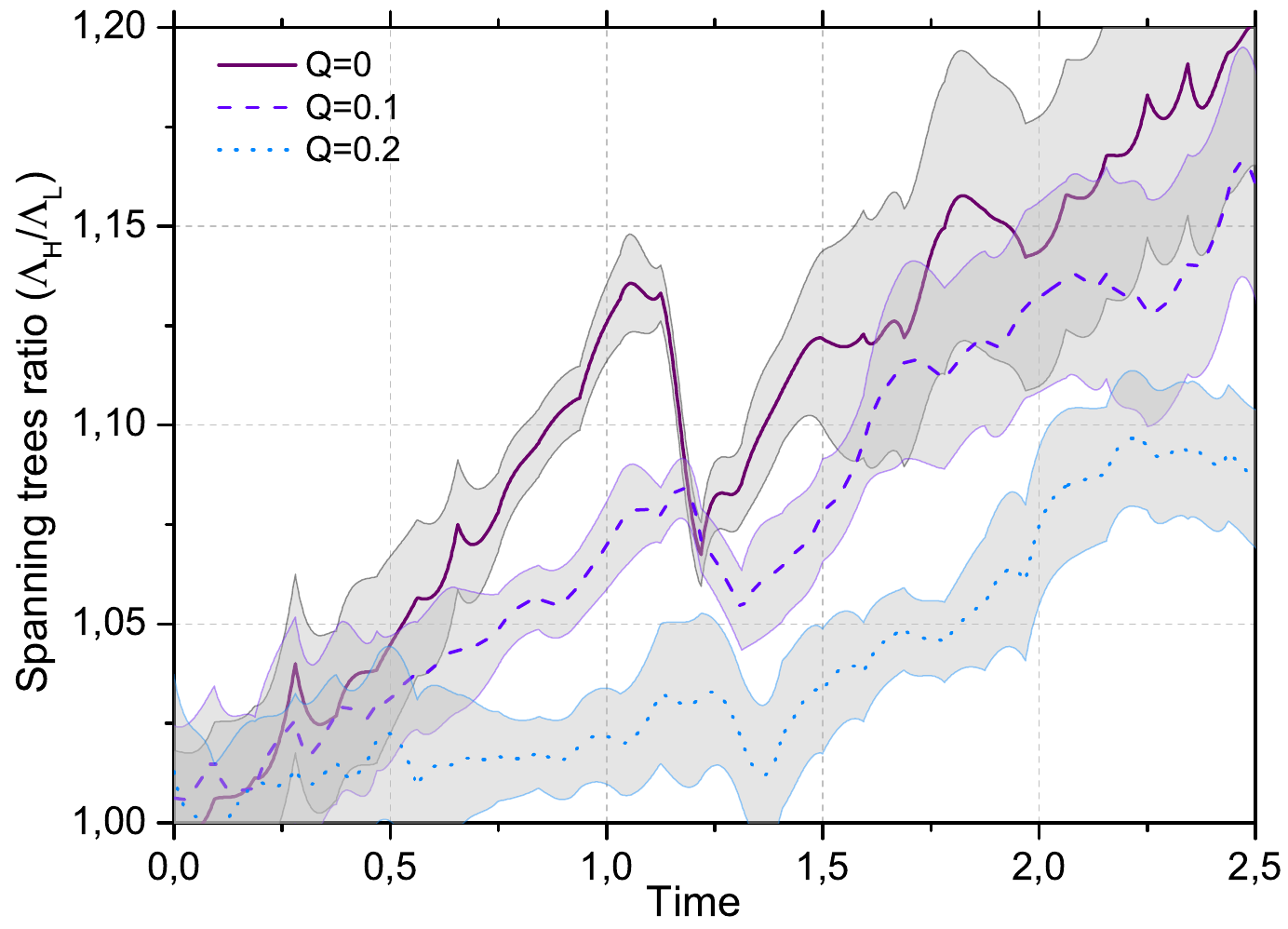}
\caption{A detail of Fig. \ref{fig:Fig1}, zooming in the interval of time between $0$ and $2.5$. }\label{fig:Fig2}
\end{figure}
Figure \ref{fig:Fig2} shows a detail of the Fig.\ref{fig:Fig1}; in particular, it refers to the initial 
evolution of the system from $t=0$ to just after the bounce. This figure shows a rapid increase of mass segregation up to the bounce time ($t\simeq 1$) which is slightly delayed by increasing $Q$. Subsequently, an inverse trend is observed for a brief time which is more pronounced for cooler initial conditions (i.e. larger compressions before the bounce, with a radius shrink of a factor 10 for $Q=0$ and $N=1024$) and, finally, mass segregation starts again with about the same rate and efficiency, as it was before. 
The decrease of the spanning trees ratio after bounce is mostly real (i.e. an actual reduction of mass segregation due to a high velocity dispersion) and only marginally due to this particular segregation indicator.

\begin{figure}
\centering
\includegraphics[scale=0.28]{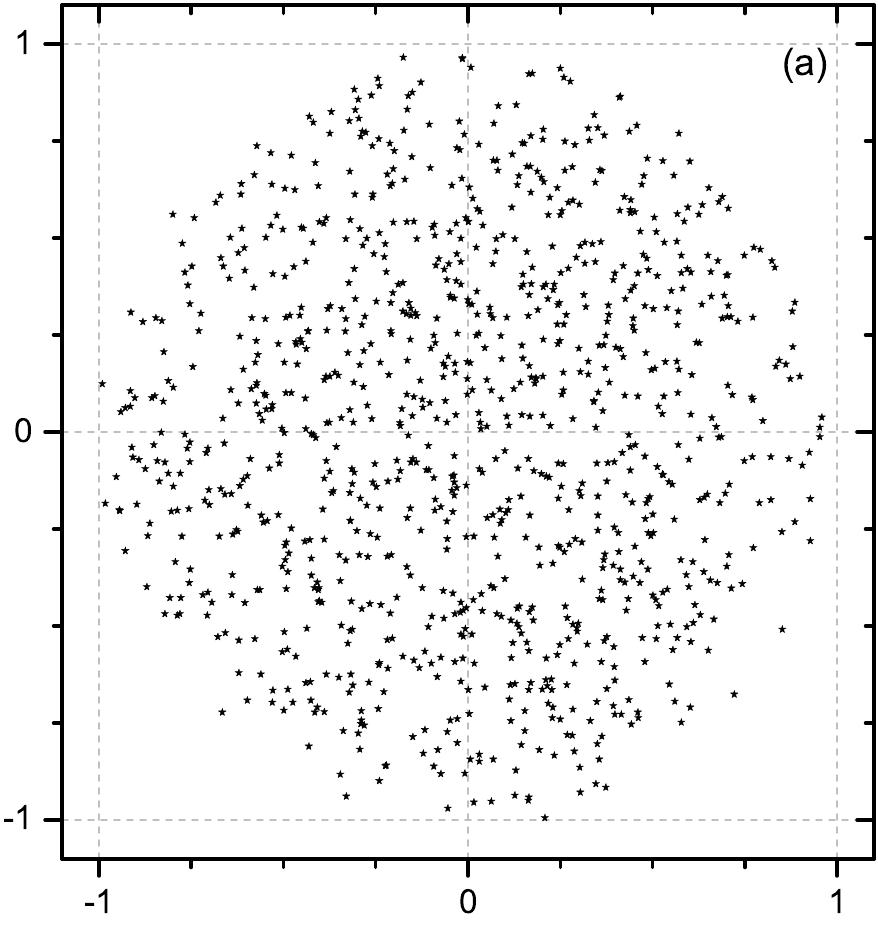}
\includegraphics[scale=0.28]{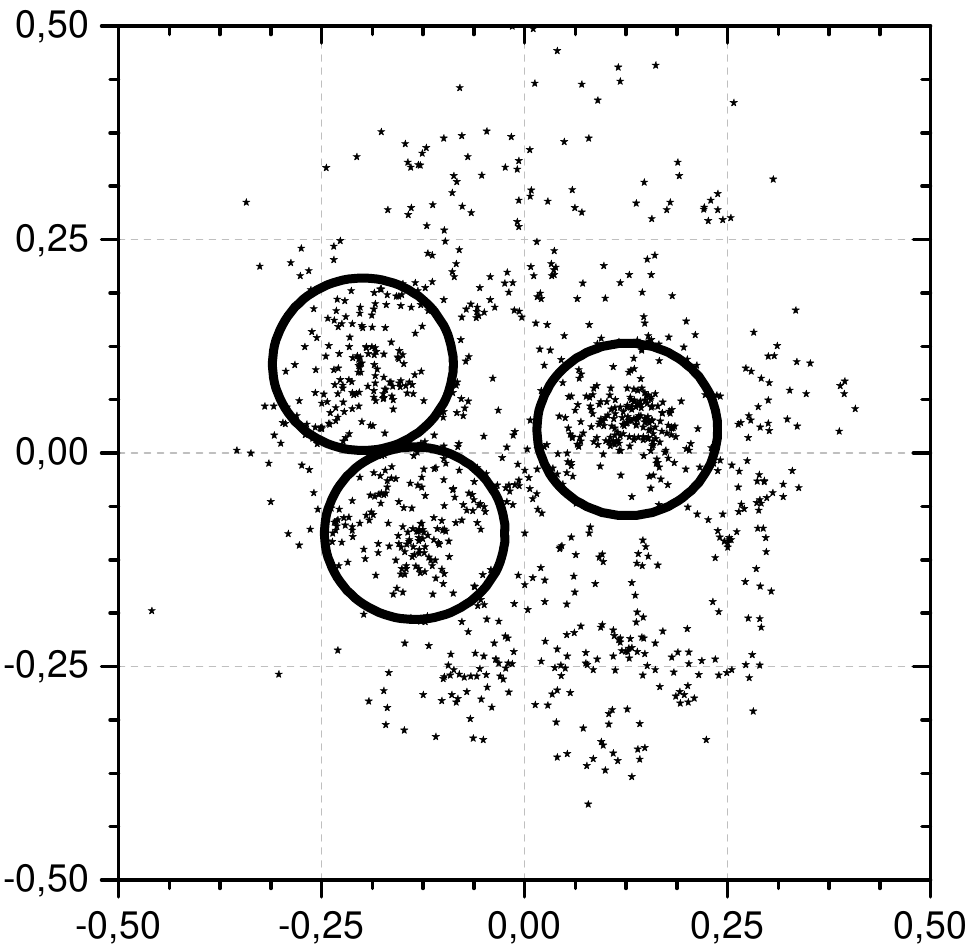}\\
\includegraphics[scale=0.28]{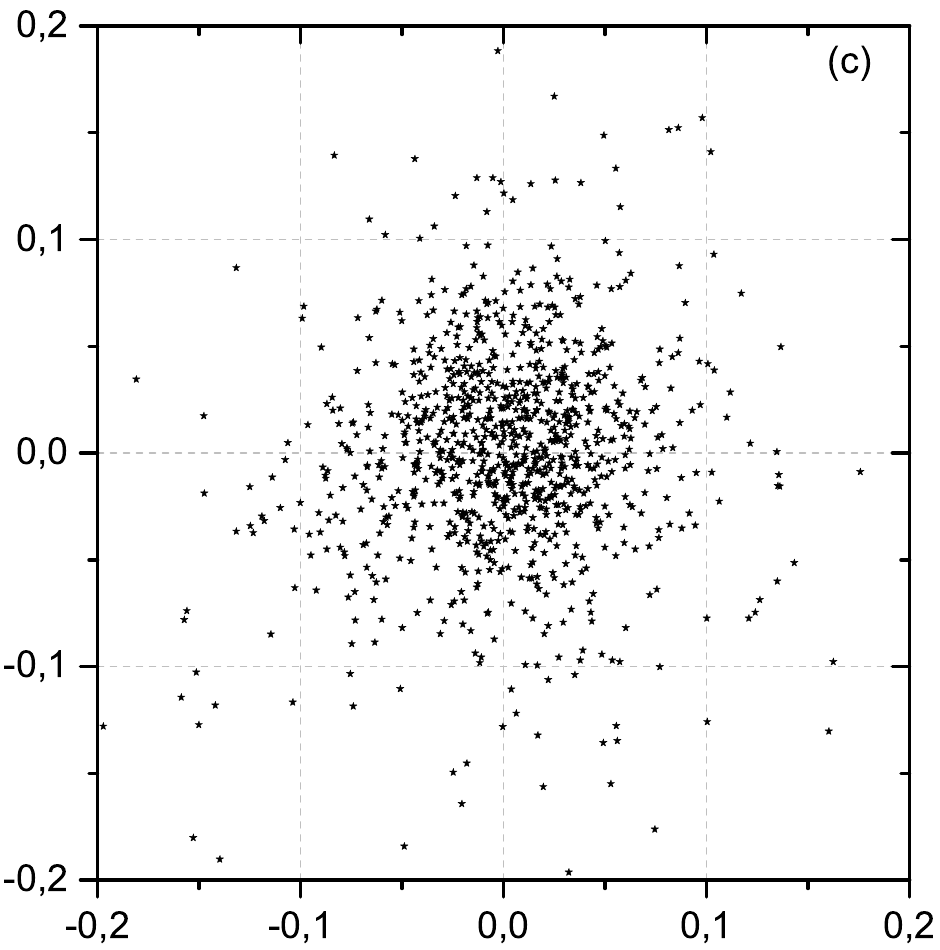}
\includegraphics[scale=0.28]{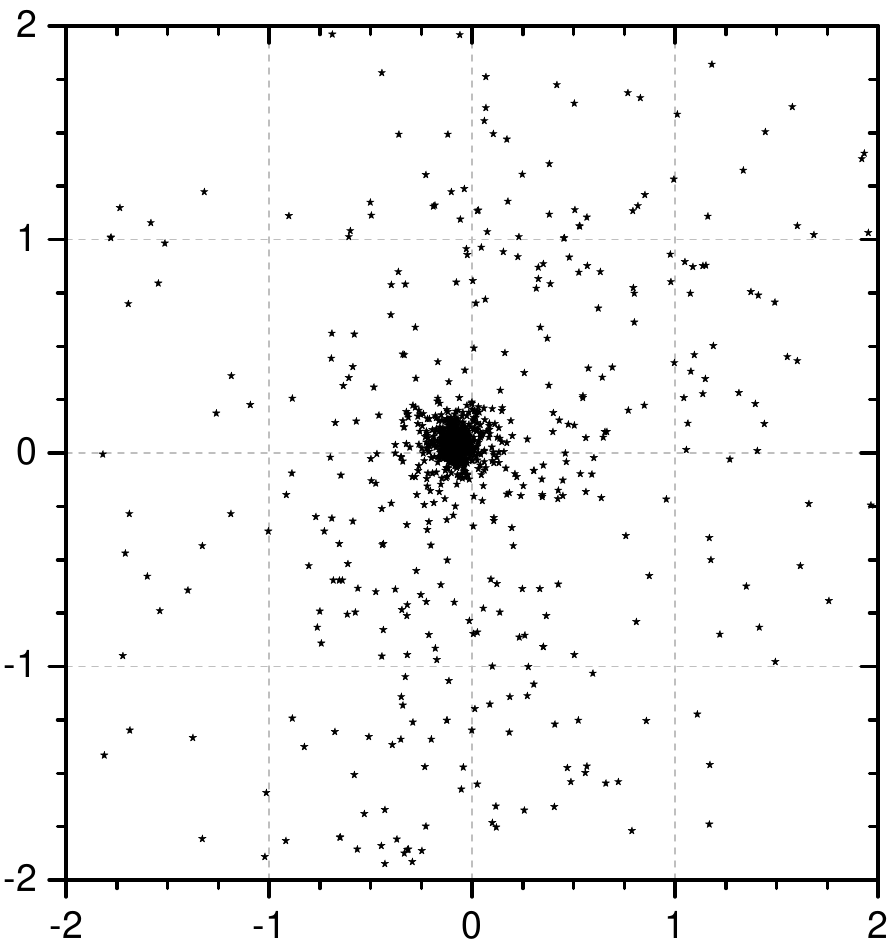}
\caption{Projection on one coordinate plane (x-y) of one of the $N=1,024$ simulated clusters with initial virial ratio $Q=0$ at four different times. Panel (a) shows the homogeneous initial distribution, panel (b) 
the formation of several sub-clumps ($t\simeq 1$), the most evident of which are circled, panel (c) refers to 
the state of maximum compression of the stellar system ($t\simeq t_B$), while panel (d) is a view of the 
cluster after the bounce ($t\simeq 2$).
Note that for display clarity we used different spatial scales for the different panels.} \label{fig:Fig3}
\end{figure}
Figure \ref{fig:Fig3} gives a visual sketch of the dynamical evolution from $t=0$ up to $t=2$ of the stellar system characterized by an initial virial ratio $Q=0$. In Fig. \ref{fig:Fig3} it is apparent that the system becomes substructured before the collapse and that all the substructures are quickly erased during the state of maximum compression.

\begin{figure}
\centering
\includegraphics[width=\columnwidth]{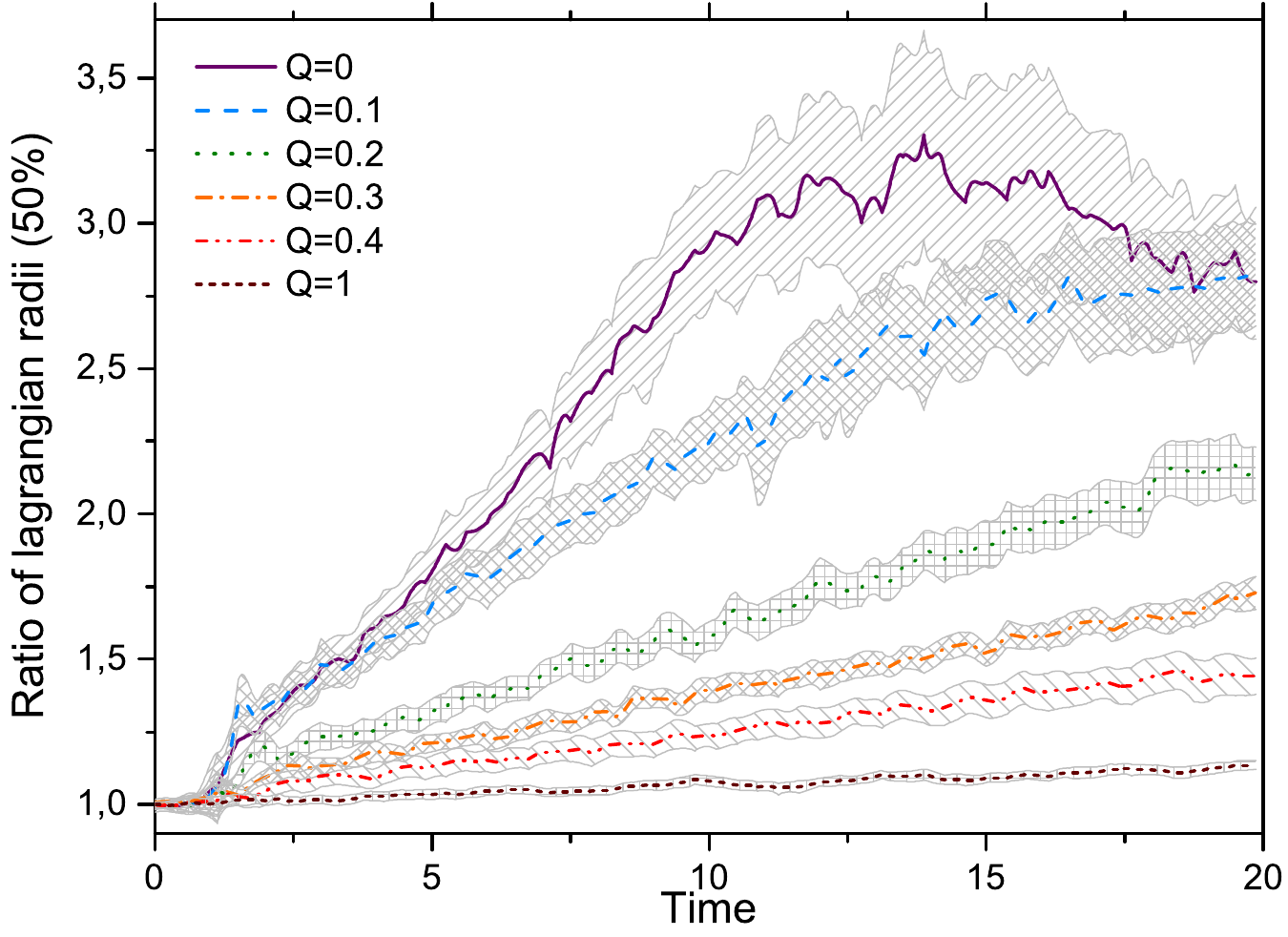}
\caption{Time evolution of the ratio of the lagrangian radius (that containing 50\% of the mass) of light to that of heavy stars in the evolving system, for the simulations of Case 1 with $N=1,024$ (see Tab. \ref{tab:tab1}). Different curves are for different values of the initial virial ratio $Q$ of the system. The pattern area represents the error associated with each curve.}\label{fig:Fig4}
\end{figure}  
For completeness, we report in Fig. \ref{fig:Fig4} the results obtained using another indicator of mass  segregation, that is the ratio between 50\% lagrangian radii of light and heavy stars.
As we said, the ratio of lagrangian radii is  as better as mass segregation indicator as the studied system maintain spherical symmetry. In our case, this is true at times greater than $\sim t_B$, where $t_B$ indicates the time at which the system reaches the state of maximum compression. In other words, we use lagrangian radii to point out the presence of a long lived mass segregation whose efficiency depends on the above
 described phenomena which occur at $t \lesssim t_B$.  
 
The phenomenon of mass segregation is particularly evident, in Fig. \ref{fig:Fig4}, for $Q=0$. On average, in 
this case, the typical size of the spatial region occupied by light stars becomes, in 6 crossing times, 
almost twice larger than the heavy particles one. On the other side, the case $Q=1$ shows an almost flat 
behaviour. This confirms again, the strong dependence of the efficiency of mass segregation upon the initial dynamical conditions of the system.

\begin{figure}
\centering
\includegraphics[width=\columnwidth]{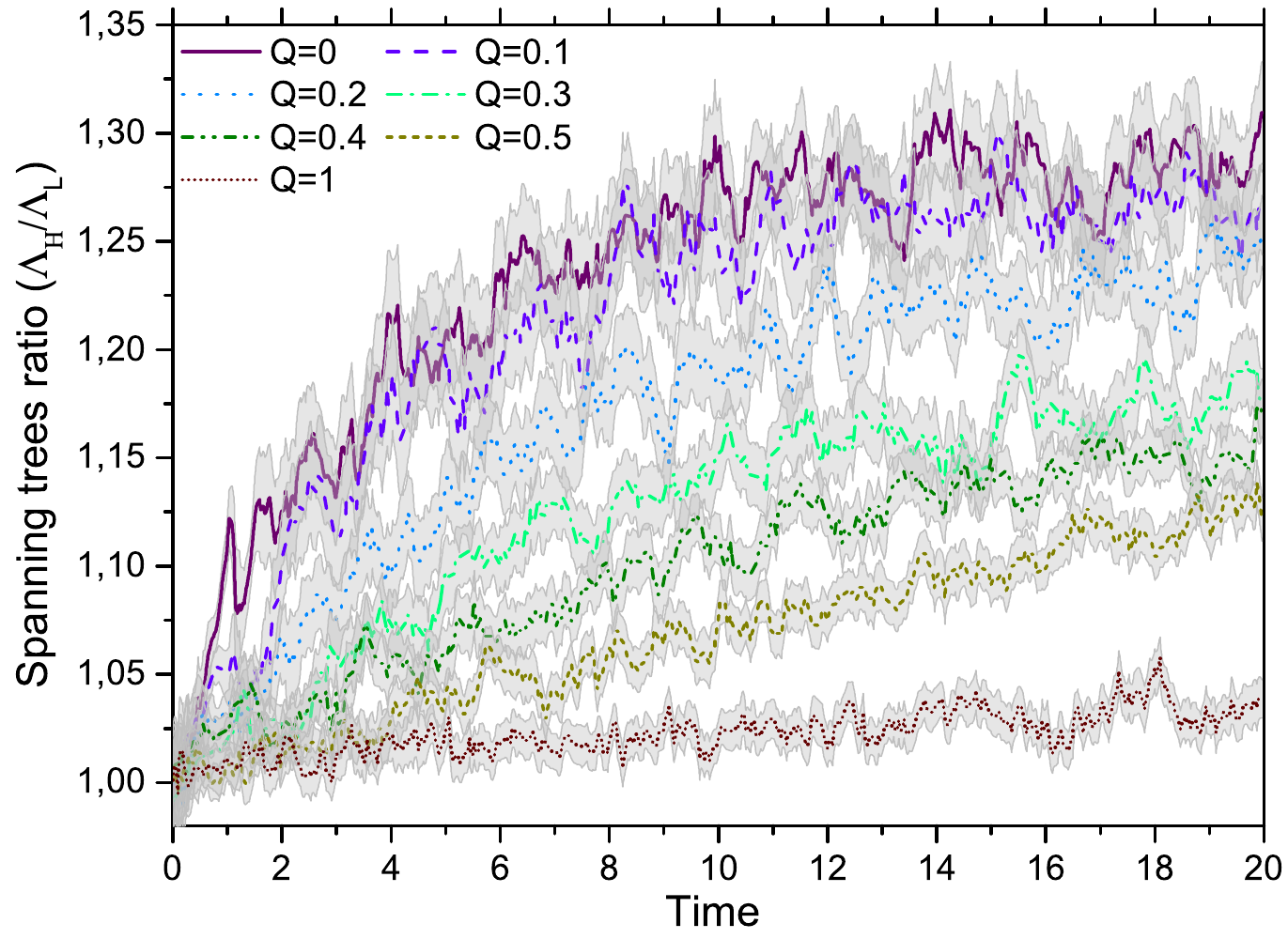}
\caption{The spanning trees ratio as a function of time, for the simulations of Case 2 with $N=1,024$ (see Tab. \ref{tab:tab1}). Values of $Q$ label the curves. Values of $Q$ 
between $0.5$ and $1$ have been studied, too, but  not displayed  here for the sake of  a more clear representation 
of the results. Each curve represents an average value of $\Lambda_h/\Lambda_l$, taking into account the 
results obtained from the single runs, while the error (standard deviation) is represented by the 
semi-transparent area.}
\label{fig:Fig5}
\end{figure}
Figure \ref{fig:Fig5} is the same as Fig. \ref{fig:Fig1} but here we include a central massive object in our simulations. The mass of the massive particle is 25 times that 
of a generic ~\lq heavy\rq~ star. The presence of this object tends to compact the curves of the spanning trees ratios with respect to the 
case shown in Fig. \ref{fig:Fig1}. It is apparent that the process of mass segregation is less efficient than in abscence of the massive object but it is still evident and strongly dependent on the violence of the collapse. The reduction of mass segregation is likely due to the stronger interaction of the massive body with the heavier stars with a consequent larger effect in spreading their orbits.

\begin{figure}
\centering
\includegraphics[width=\columnwidth]{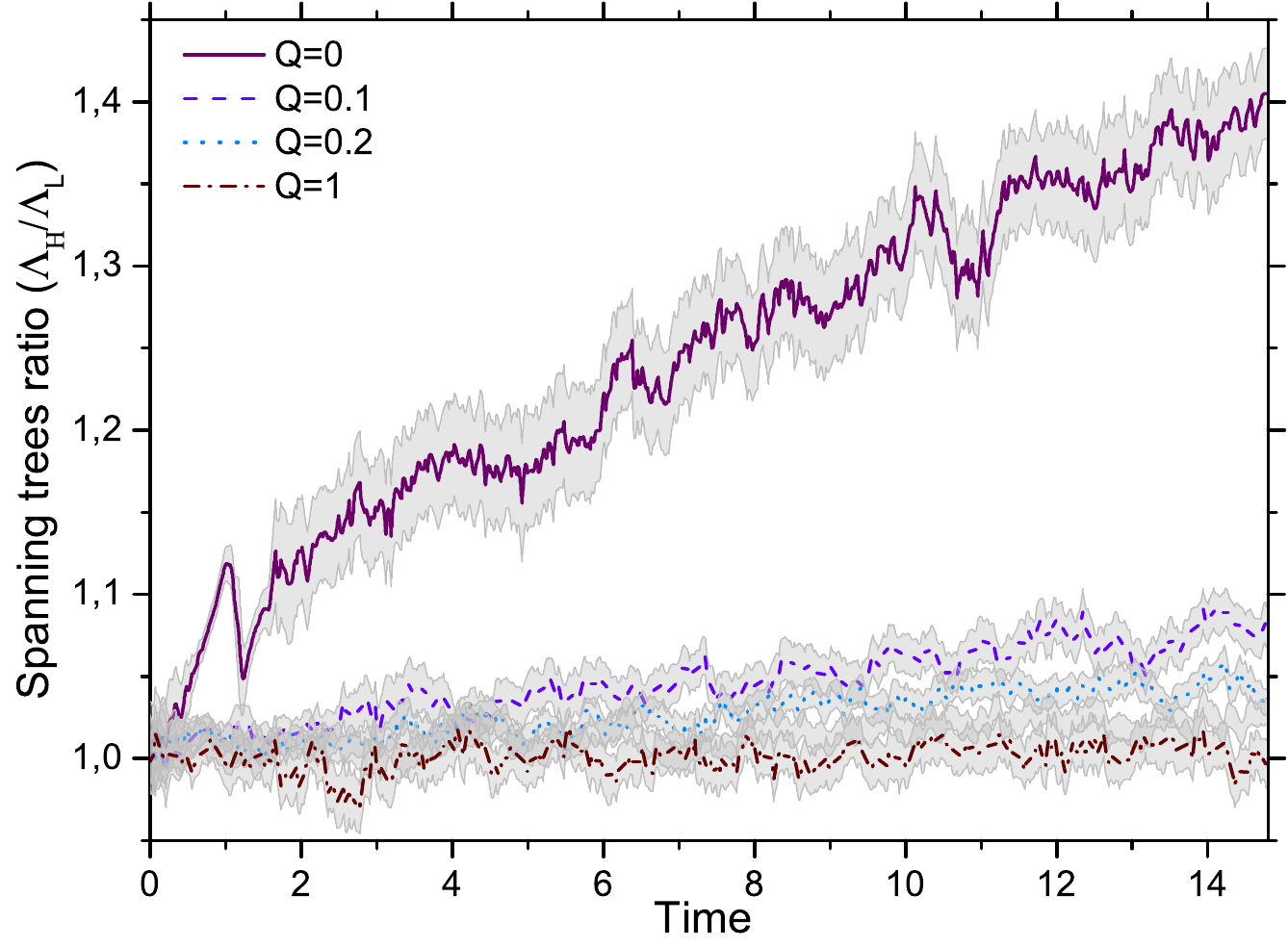}
\caption{The spanning trees ratio as a function of time, for the simulations of Case 3 with $N=1,024$ (see Tab. \ref{tab:tab1}). Values of $Q$ between $0.4$ and $0.9$, both taken into account in our simulations, 
are not shown here to have a clearer representation of the results. Each curve represents an average value of 
$\Lambda_H / \Lambda_L$, taking into account the results obtained from the single runs, while the error 
(standard deviation) is represented by the semi-transparent area.}
\label{fig:Fig6}
\end{figure}
\begin{figure}
\centering
\includegraphics[scale=0.33]{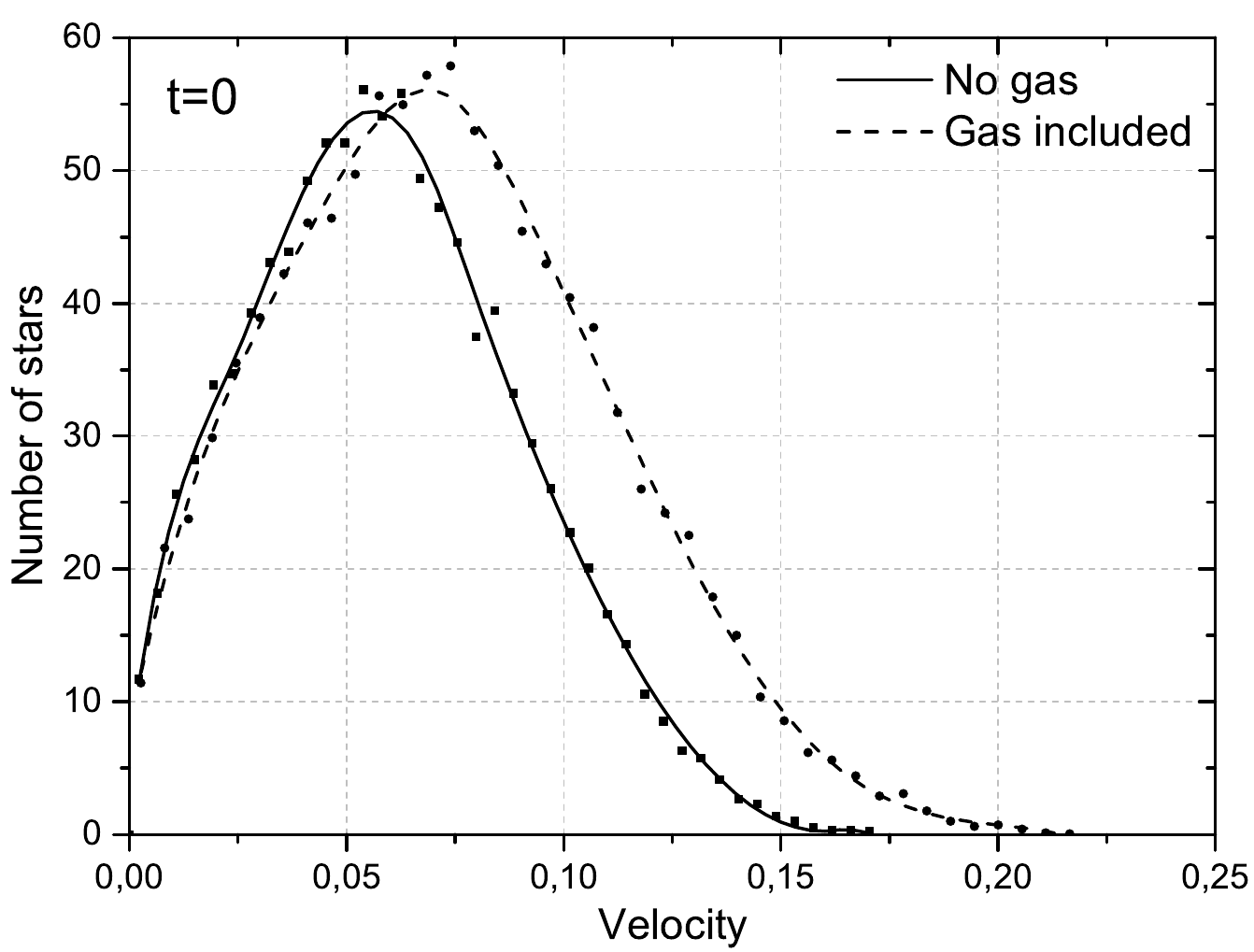}
\includegraphics[scale=0.33]{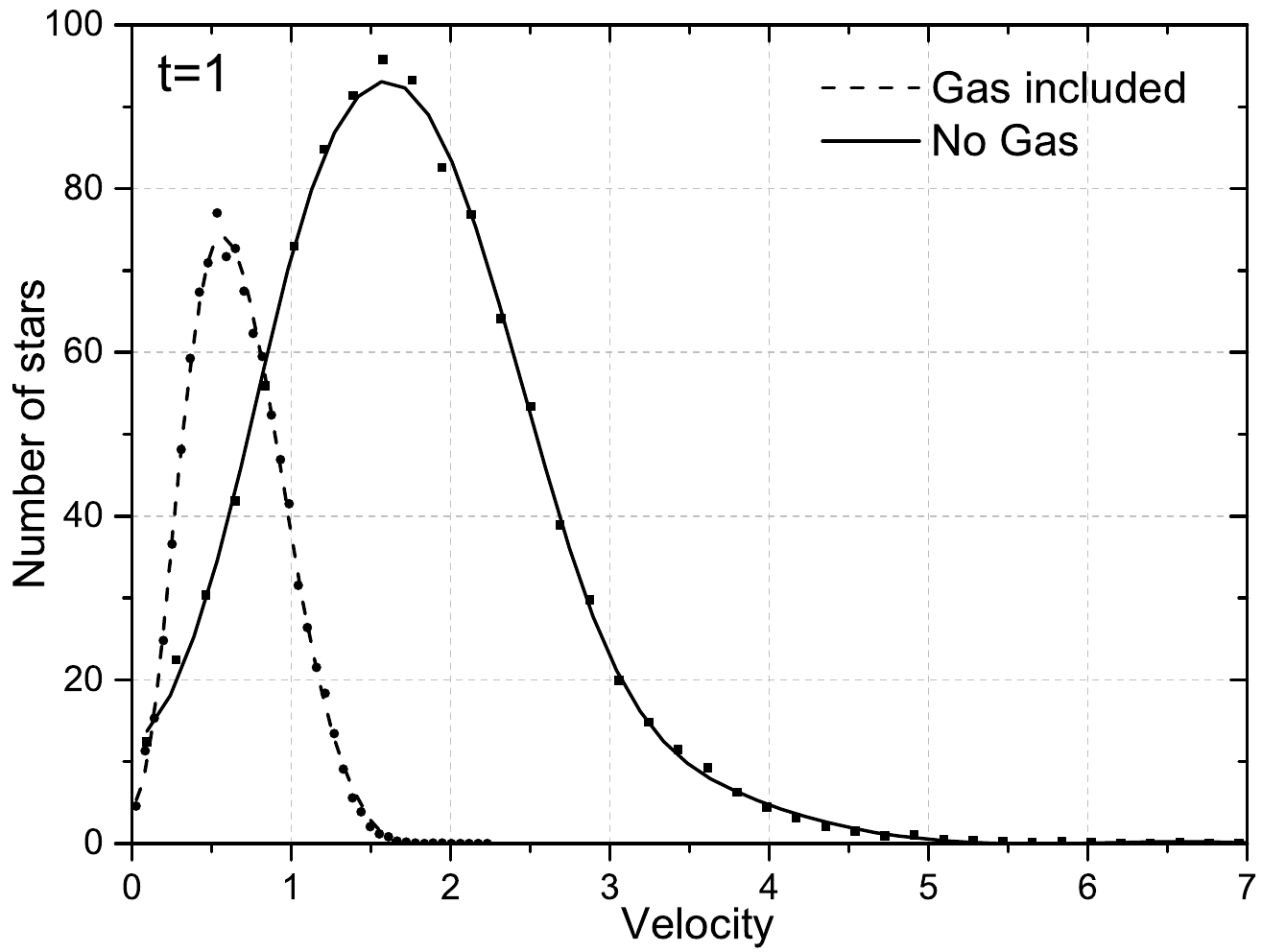}\\
\includegraphics[scale=0.33]{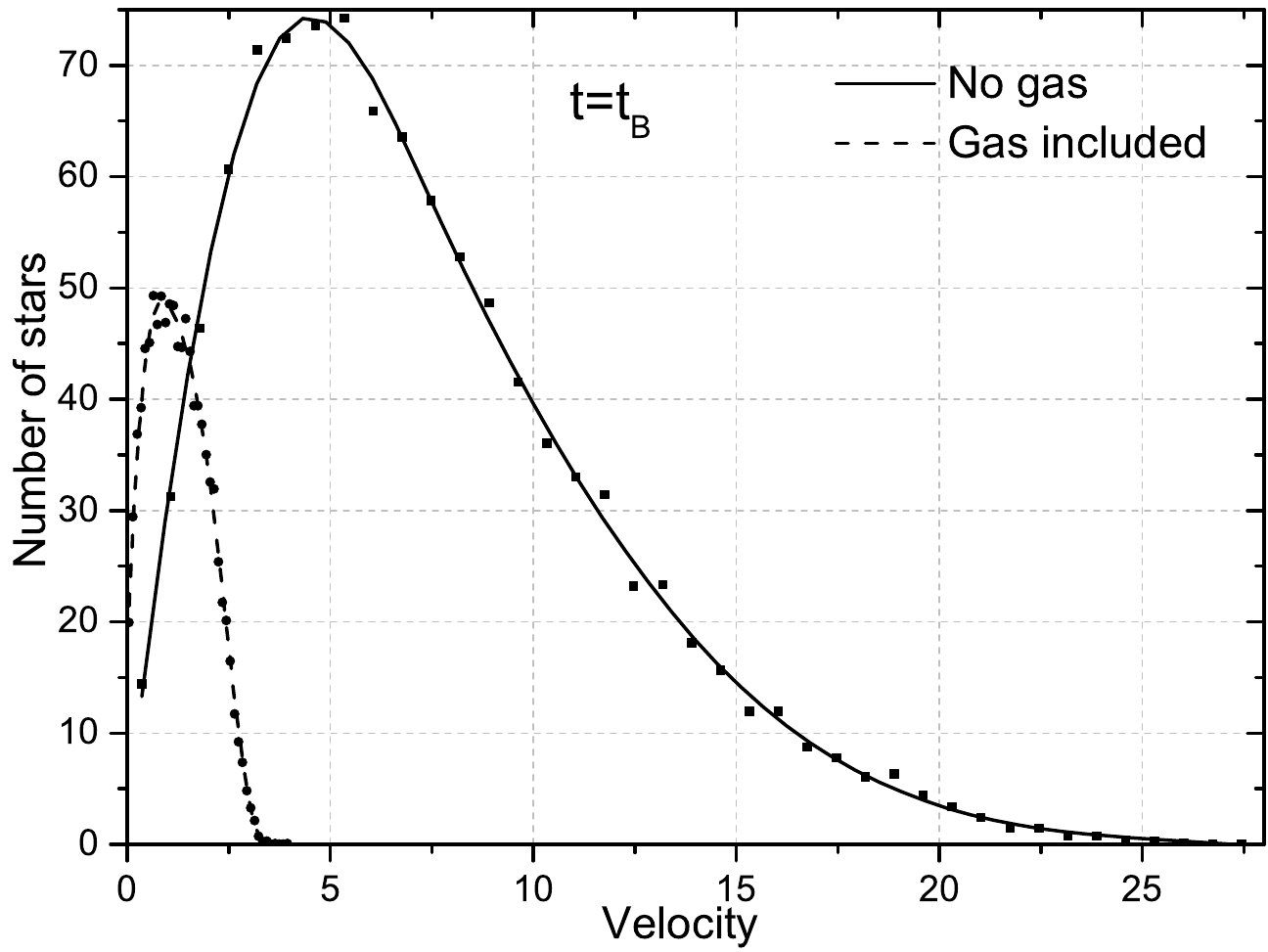}
\includegraphics[scale=0.33]{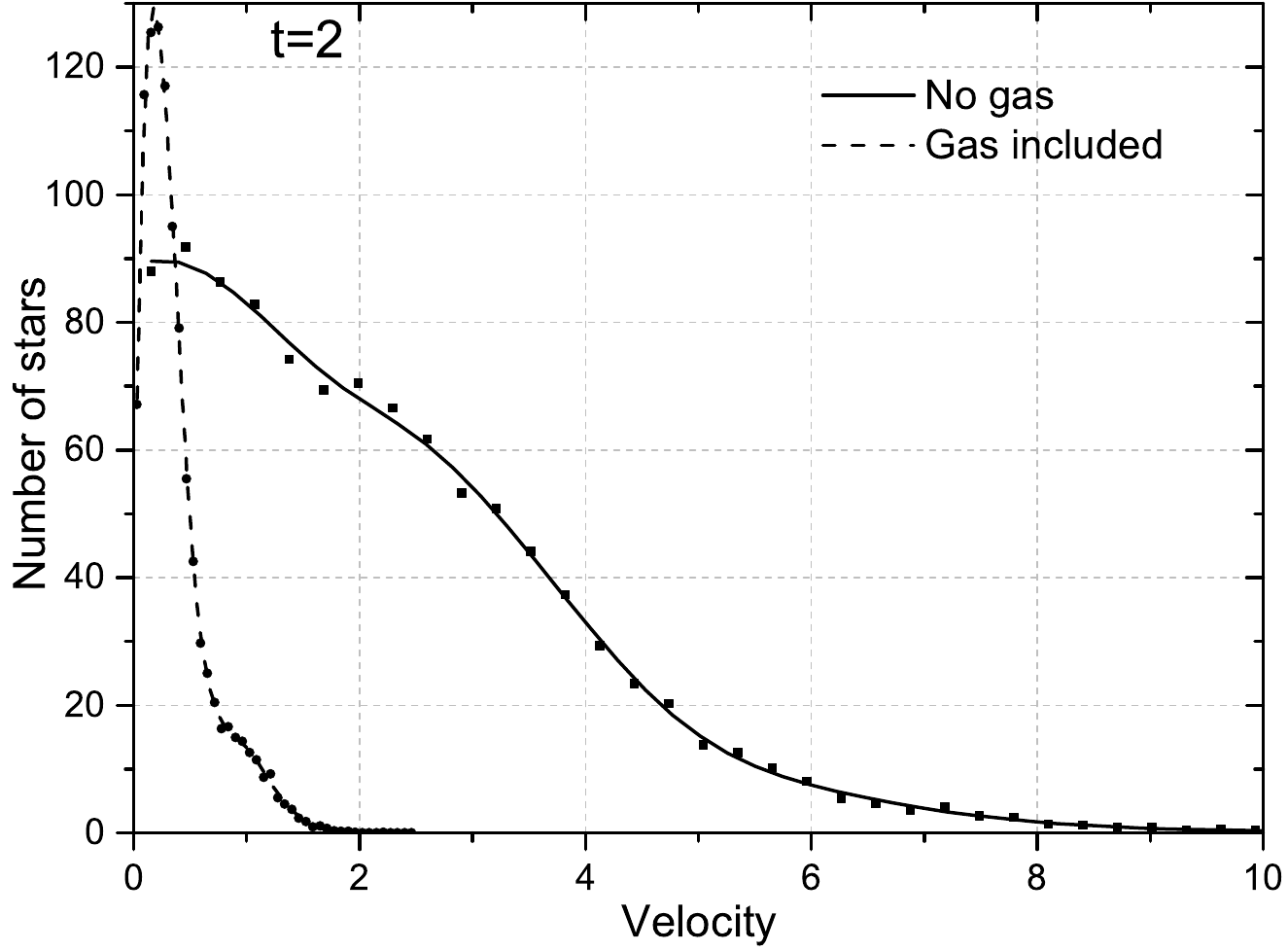}
\caption{The star velocity distribution in the case of of an $N=1,024$  system without any background gas (Case 1, black triangles fitted by a solid line), and in presence of a background static gas (Case 3, black squares, fitted by a dashed line). The initial value of $Q$ is 0.1. The four panels represent different times: the initial state, the situation after 1 crossing time, at the time of the bounce and after 2 crossing times.  Velocities are in code units, as specified in Sec. \ref{sec:model}.} 
\label{fig:Fig7}
\end{figure}

 
Figure \ref{fig:Fig6} is the analogous of Fig. \ref{fig:Fig1} with the inclusion of a stationary gaseous 
background whose gravitational potential is given by Eq. \ref{eq:Plu} with  $a(t)=a(0)=R$ and assuming a star formation efficiency $\mathcal S=0.3$. The inclusion of a background gravitational potential let us take into account that the majority of young and very young star clusters are still embedded in their proto-cloud. This residual gas, when sufficiently abundant, acts as a background potential which can affect the dynamical evolution of the stellar cluster. In this work we modelled the presence of a gas as an analytical potential to add to the pair gravitational interaction  of the bodies in the system. This  does not represent very correctly the real astrophysical situation but, using our simple model, we can at least account at order of magnitude accuracy the gaseous phase contribution. The most evident effect of the addition of the analytic gravitational potential is smoothing out the 2-body encounters, decreasing the efficiency of mass segregation on both small and longer time scales. This smoothing is due to that the inclusion of a significant quantity of regular potential reduces stochasticity of the star trajectories and so the role of 2 body scattering and so enlarge the time to equipartition. 
 As we can see in Fig. \ref{fig:Fig6}, the results for the case of $Q=0$ are very similar to those obtained for the same 
case but without the presence of gas; it is still evident the formation of sub-clumps and a quick mass 
segregation, before the bounce, as much the inverse trend of the spanning trees ratio around the time 
corresponding to the  maximum compression of the system. The mass segregation on a longer time scale is 
slightly less than that observed for the same value $Q=0$ in Fig. \ref{fig:Fig1}. For values of the initial 
virial ratio  $Q \gtrsim 0.1$ the situation deeply changes. The presence of gas reduces the role of 
close encounters between stars and the process of mass segregation is less efficient. Although this point 
deserves 
a deeper investigation, it is possible to point out that the case $Q=0.1$ is profoundly different respect to 
the 
case of absence of gas as it is seen also in the evolution of the velocity distribution of the 
stars (Fig. \ref{fig:Fig7}).

Figure \ref{fig:Fig7} shows the distribution of the velocities of the stars at different times. In both the 
cases of presence and of absence of a gas component, the initial velocity distributions follow 
approximatively the same trend. The situation deeply changes after $\sim$1 crossing time: the velocity 
dispersion in the system which does not include gas becomes rapidly broader than the other case. This implies that the background potential reduces the efficiency of the 2-body interactions whose effects on a reduced mass segregation we see in Fig. \ref{fig:Fig7}  and, then, reduces the 
possibility to form gravity-driven substructures (clumps) and the efficiency of mass segregation. At the 
bounce, the difference between the two systems is even more marked and, after $\sim$2 crossing times, the 
two distributions are completely different. In particular, the presence of the gas tends to pack the velocity 
distribution toward the same, small, velocity. On the other hand, the velocity distribution, when no gas is 
present, is extended over a huge range of velocities. This is the natural consequence of more efficient 
2-body encounters which make the system segregate masses more rapidly.

We have also studied the case of an expanding background gas, represented as a Plummer sphere where the length scale, $a$, is not constant but, rather, evolves according to the law $a(t)=(0)\exp(t/\tau)$, where 
the time scale $\tau$ is set equal to the crossing time of the system ($t_c$) and  $a(0)=R$. The total 
amount of the mass in gas is given by the star formation efficiency parameter $ $ (see Eq. 
\ref{eq:sfe}); we recall that, in our simulations, we have $\mathcal S=0.3$.
The law is such that the core radius amplify its value of a factor $k$ after a time $t_k=\tau \ln k$.

\begin{figure}
	\centering
	\includegraphics[width=\columnwidth]{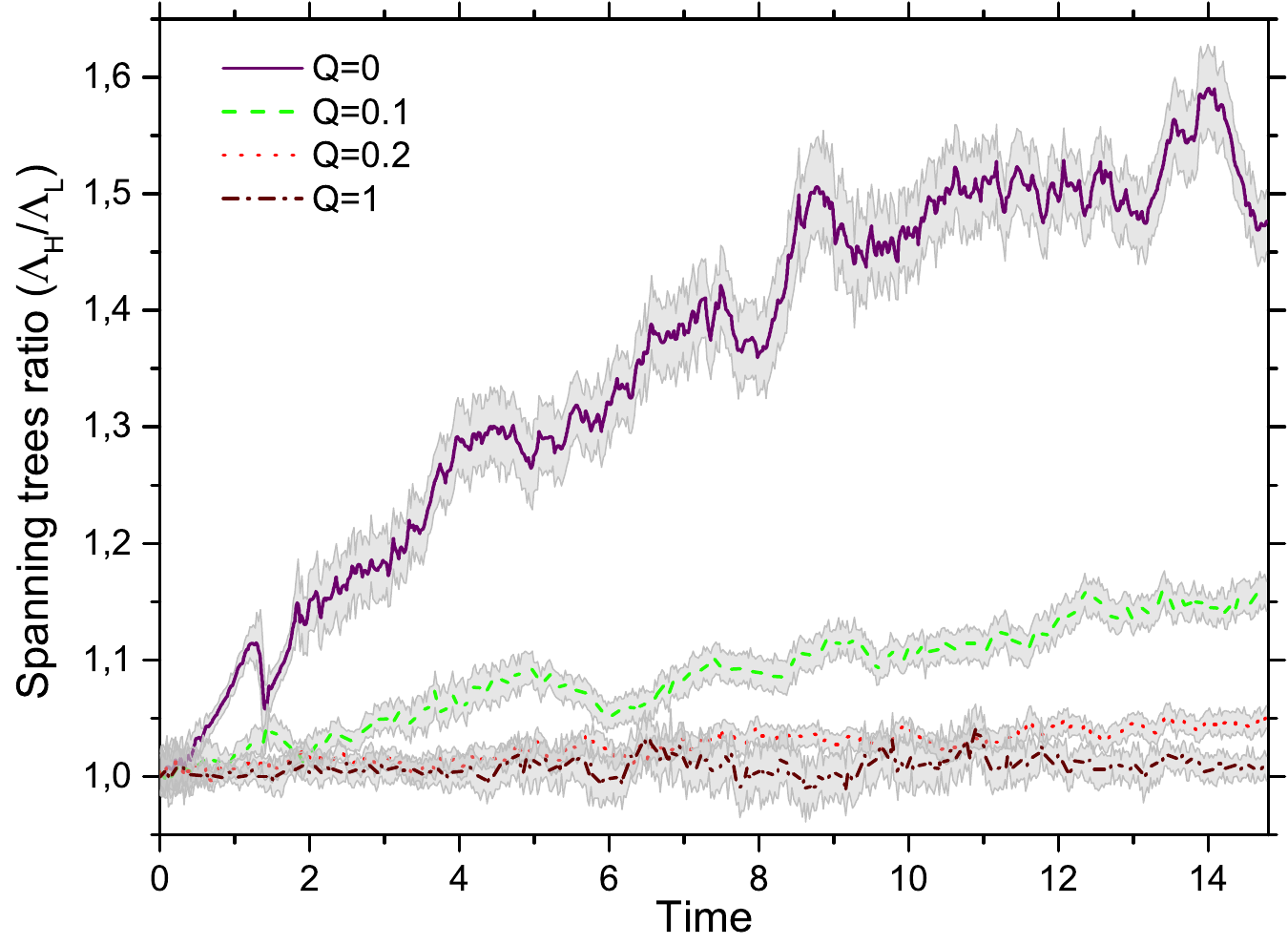}
	\caption{The spanning trees ratio as a function of time, for the simulations of Case 4 with $N=1,024$ (see Tab. \ref{tab:tab1}). The different values of the initial $Q$ 
	are labelled. Each curve represents an average value of $\Lambda_H / \Lambda_L$, taking into account the 
	results obtained from the single runs, while the error (standard deviation) is represented by the 
	semi-transparent area.}
	\label{fig:Fig8}
\end{figure}
Figure \ref{fig:Fig8} gives the spanning trees ratio in this expanding case, for 4 values of the initial virial ratio $Q$.
We note how the spanning trees ratio do not show significant differences from the behaviour they have in the static gas potential case (Fig. \ref{fig:Fig6}). In particular, the curves in Fig. \ref{fig:Fig8} reproduce the trends reported in Fig. \ref{fig:Fig6} (the static gas case) for values  $0.1 \leq Q \leq 
1$. Compared to the static gas case, here the 
spanning trees ratio are comparatively larger, stabilizing at a value of $\sim 1.5$ after $\sim 8t_c$, in the 
zero initial velocities case ($Q=0$). This means that, also in this case, the violent mass segregation effect 
is quite evident.

\subsection{Dependence on $N$ and  on the heavy-to-light stars number ratio}
\label{s3.1}
\begin{figure}
	\centering
	\includegraphics[width=\columnwidth]{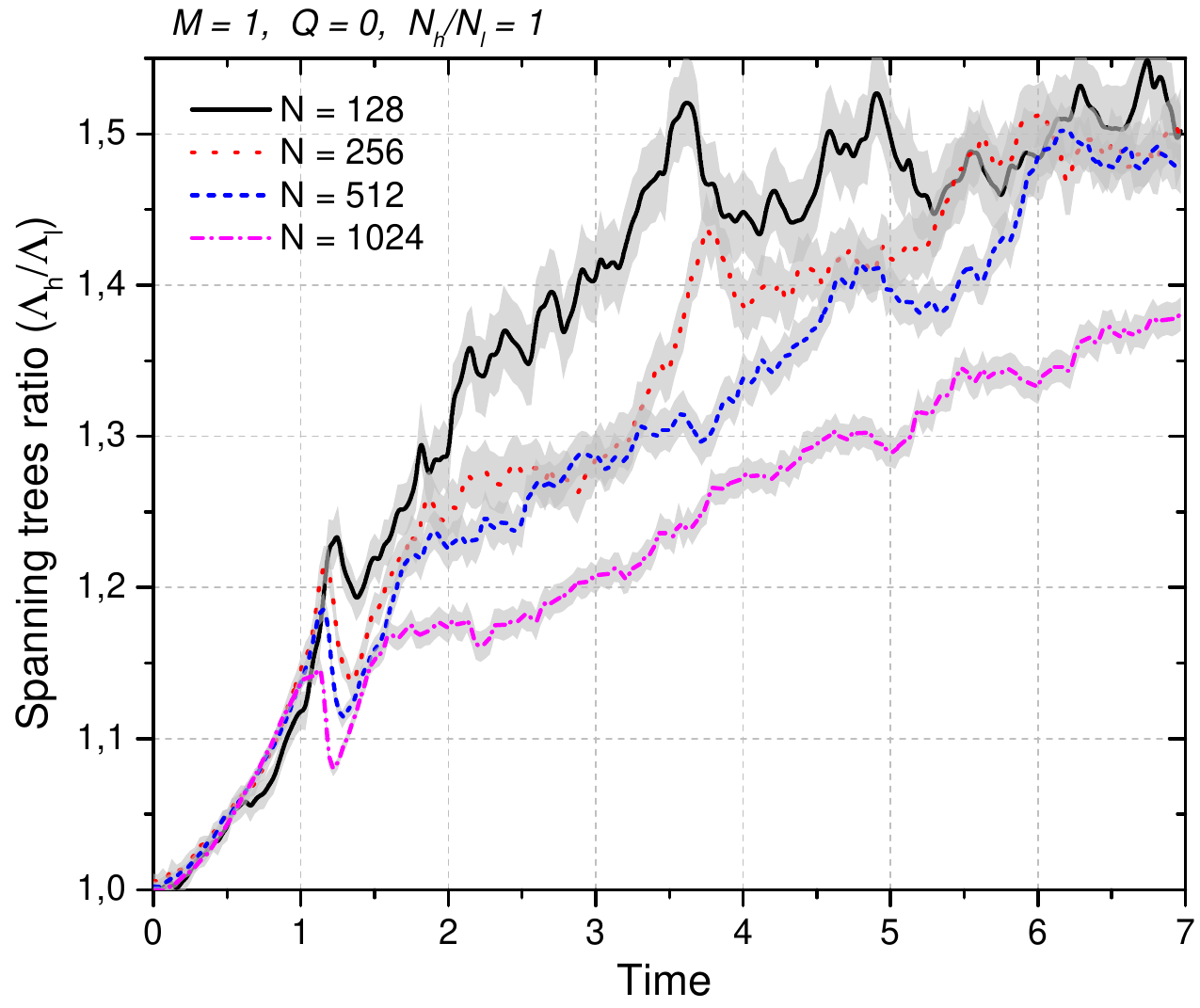}
	\caption{The ratio between $\Lambda_h$ and $\Lambda_l$, for different values of $N$, as a function of time. Each curve represents an average value of $\Lambda_h / \Lambda_l$, obtained from 50 runs, and the error (standard deviation) is represented by the  grey, semi-transparent area.}
\label{fig:Fig9}
\end{figure}
Figure \ref{fig:Fig9} (which refers to an initial $Q=0$ and same number of heavy and light stars over the range $128 \leq N \leq 1024$) is very useful to show how the segregation process is efficient on the whole range of $N$ investigated, with greater efficiency for less abundant systems. This means that the mass segregation is driven by violent relaxation as coarse-grain main engine and `tuned' by the fine-grain, speeded up, two-body interaction whose amplitude is larger for smaller $N$.
Actually, the fast growth (in units of initial crossing times) of the mass segregation after the system bounce in the cases of lower $N$ is interpreted in terms of the small ratio of the half mass 2-body relaxation time to the crossing time \citep{binney2008}

\begin{equation}
\frac{t_{rel}}{t_{c}}=\frac{1}{10} \frac{N}{\ln N},
\end{equation}

which ranges in $2.64 \leq t_{re}/t{cr} \leq 14.8$ for $128\leq N \leq 1024$.
Additionally, formation of clumps during the pre-bounce phase speeds up their mass segregation thus building up a larger mass segregation at the bounce time.
We have also performed some simulations varying the ratio of the number of heavy to light stars in a $N=1,024$ stars system, starting from cold ($Q=0$) conditions. This was done to be more representative of a realistic mass distribution, which in real systems is usually a decreasing function of the mass, corresponding to $N_h/N_l < 1$. 
\begin{figure}
	\centering
	\includegraphics[width=\columnwidth]{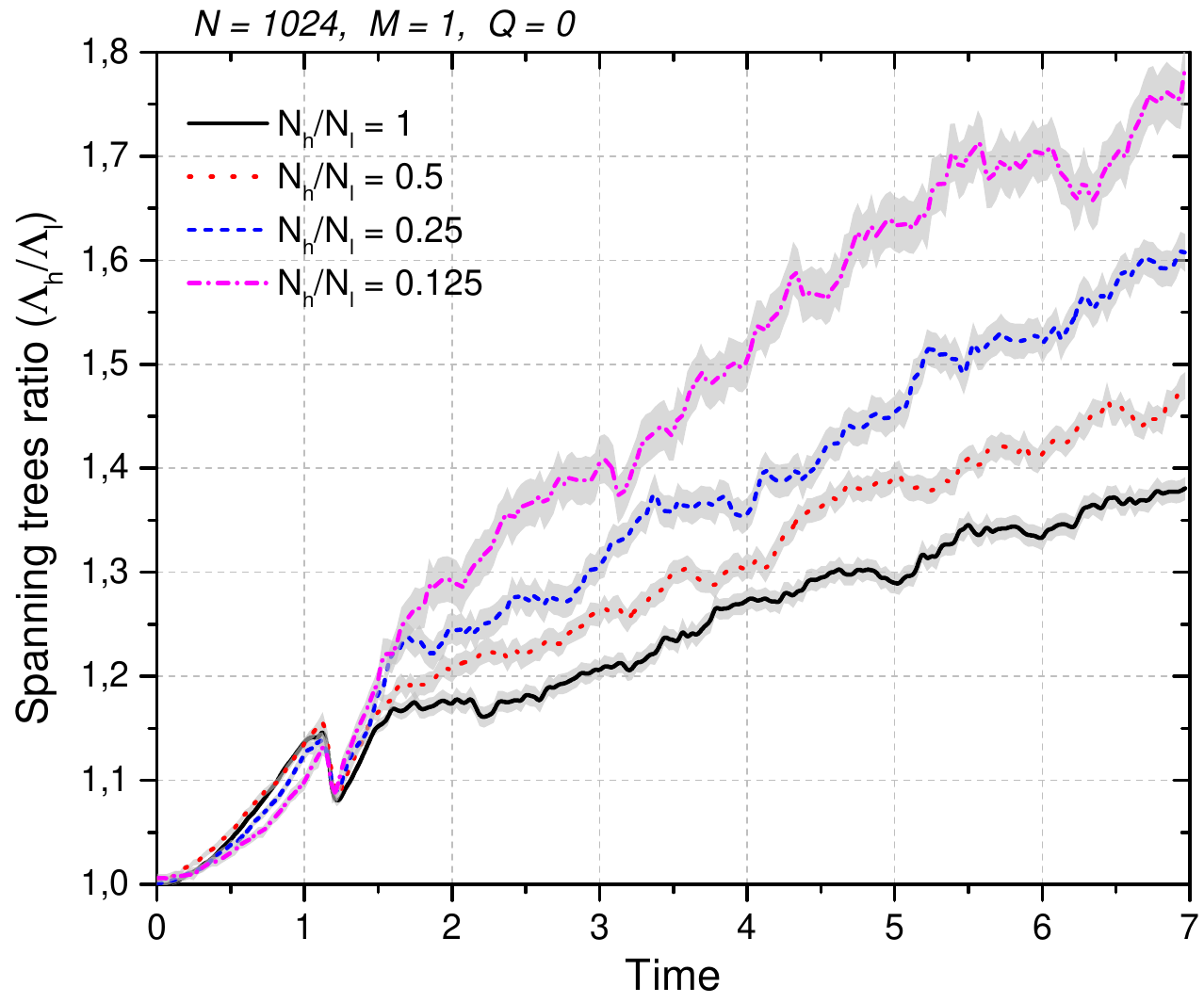}
	\caption{The ratio between $\Lambda_h$ and $\Lambda_l$, for different values of the ratio $N_h/N_l$, as a function of time (simulations of Case 5 with $N=1,024$ (see Tab. \ref{tab:tab1})). Each curve represents an average value of $\Lambda_h / \Lambda_l$, obtained from 50 runs, and the error (standard deviation) is represented by the  grey, semi-transparent area.}
\label{fig:Fig10}
\end{figure}
The result, in terms of spanning trees ratio as function of time is reported in Fig. \ref{fig:Fig10}. We notice an increasing evidence of mass segregation at decreasing ratio of heavy to light stars abundance, thing interpreted as due to the expected scaling of the segregation time with $m_h/\sqrt{\langle m\rangle}$ where the average mass $\langle m\rangle$ obviously reduces at decreasing $N_h/N_l$.

\section{Discussion}
\label{sec:disc}
As we discussed in Sect. \ref{sec:results}, the spanning trees curves represented in Fig. \ref{fig:Fig1} and, in more detail, in Fig. \ref{fig:Fig2} (referring to a system with $N=1,024$, without gas and without a central massive object) show a rapid increase before the system collapses ($t\lesssim t_B$), an inverse trend around the state of maximum compression ($t\simeq t_B$) to increase again after the collapse ($t > t_B$). The rapid increase of mass segregation before the bounce is that observed in the substructures (clumps) that form during the first system collapse. When the clump merging process occurs, the mass segregation is only partially preserved, and Fig.\ref{fig:Fig2} shows an inverse trend, indeed. After the merging process, mass segregation continues with approximatively the same efficiency as before the bounce but, this time, no sub-structures appear; therefore, mass segregation continues inside the dense core formed just after the bounce. It is thus clear that clumps are not the unique cause of the observed mass segregation on short time scales, and the same can be said for the very dense core. Actually, both the two phenomena play a role because the 
first one acts on a very short time-scales, $t<1$, while the second one is responsible of the long-lived mass segregation for $t\gtrsim t_B$. Around the collapse phase ($t\simeq t_B$) a transition between the two regimes occurs. At $t\simeq t_B$ the substructures formed during the initial phase are rapidly lost and a small, compact, core forms. 

\begin{figure}
	\centering
	\includegraphics[width=\columnwidth]{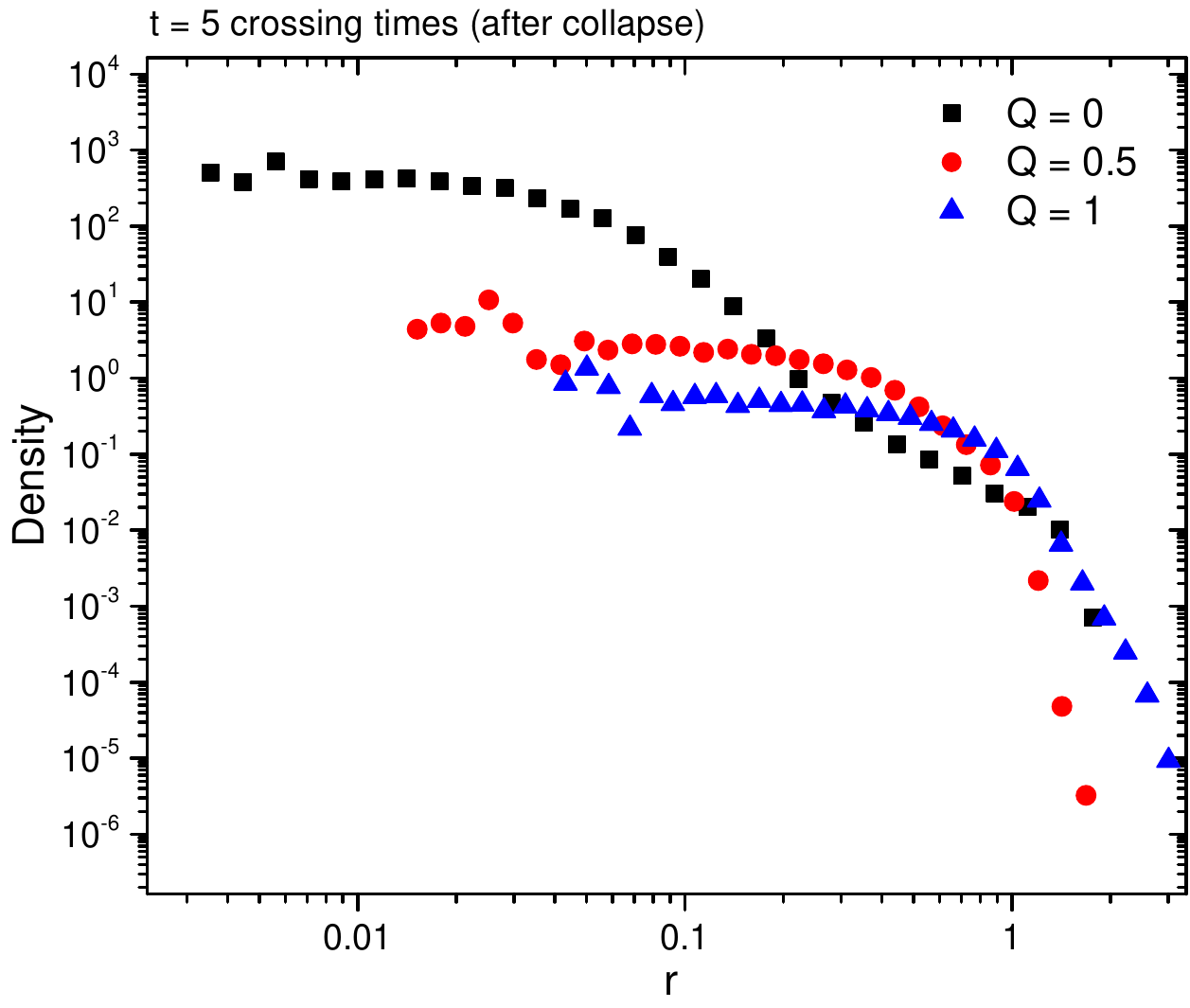}
	\caption{Density profiles of the systems of the simulations of Case 1 with $N=1,204$ and with $Q=0$ (black squares), $Q=0.5$ (red circles) and $Q=1.0$ (blue triangles) at $5$ crossing times after the collapse. The crossing time here is that of the dense system formed after the collapse.}
	\label{fig:Fig11}
\end{figure}

This situation is evident for ~\lq cool\rq~ systems ($Q\lesssim 0.3$); for $Q\gtrsim 0.3$, sub-clumps do not form at all, although a degree of mass segregation significantly 
greater than that in the equilibrium case ($Q=1$) is established. In these ~\lq warm\rq~ cases ($Q\gtrsim 0.3$), the only mechanism active in both the rapid and the secular mass segregation is the dynamical evolution of the dense core formed after the collapse. Nevertheless, the differences we see in the spanning trees for $t > t_B$ are due that, as already evidenced by \citet{allison2009}, initially ~\lq cold\rq~ and clumpy stellar systems are able to collapse in a smaller and denser system than in warm and smooth clusters (see, for example, Fig. \ref{fig:Fig11}). This implies that, after the collapse, the emerging dense system evolves following a shorter dynamical time scales. 
 \begin{figure}
 	\centering
 	\includegraphics[width=\columnwidth]{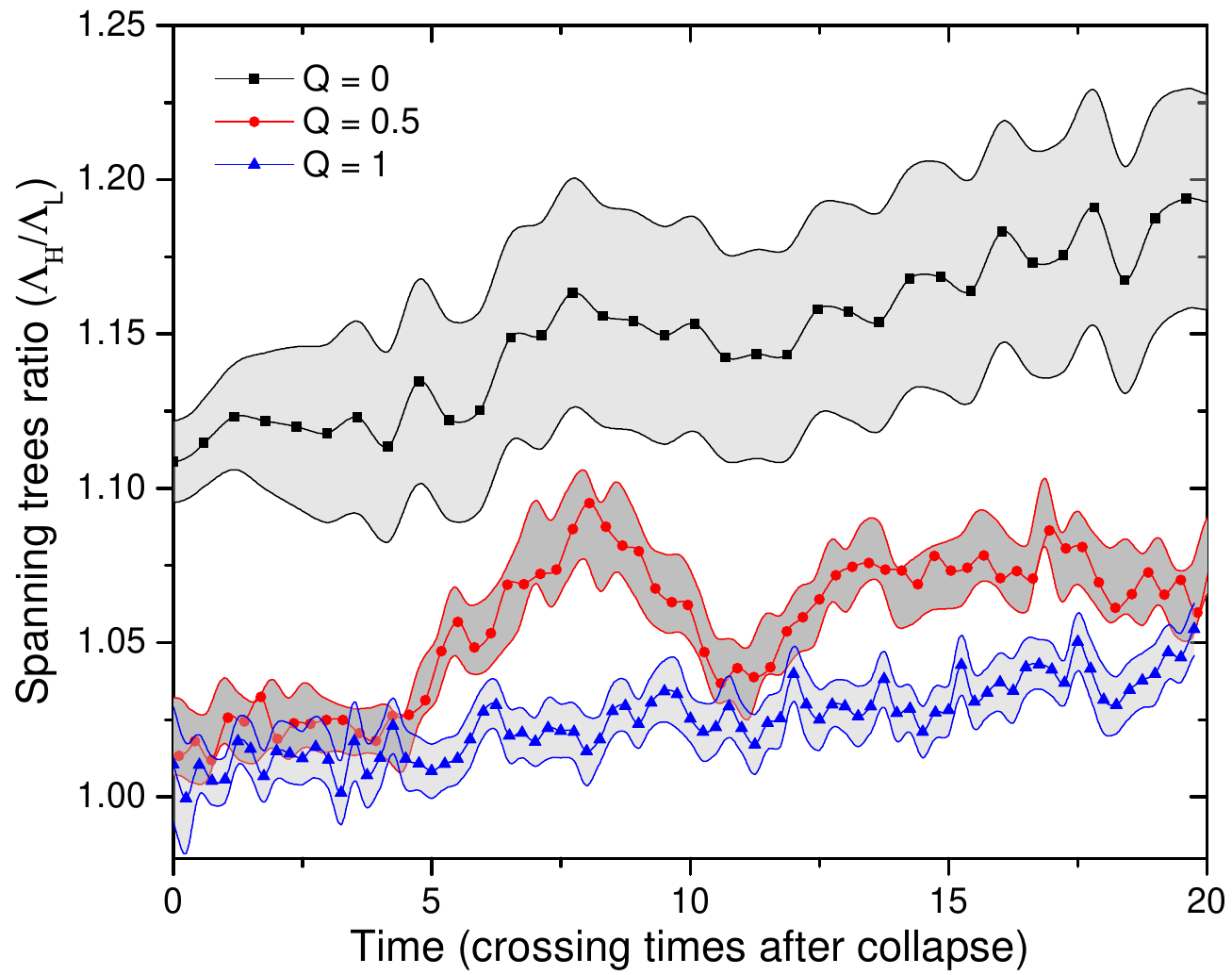}
 	\caption{Spanning trees ratio of the stellar systems of the simulations of Case 1 with $N=1,024$ and initial virial ratios $Q=0$ (black line with squares), $Q=0.5$ (red line with circles) and $Q=1.0$ (blue line with triangles) as a function of time. Time here is expressed in units of the crossing time of the dense virialized system that forms after the collapse.}
 	\label{fig:Fig12}
 \end{figure}
 In particular, Fig. \ref{fig:Fig12} shows the spanning trees ratios as a function of time expressed in units of the crossing time of the virialized compact system that forms after collapse. The curves cover the interval from $0$ to sbout $3.3$ in the time units of Fig. \ref{fig:Fig1}, which are the initial crossing times of the examined systems.
 Notice that the curves are quite similar in terms of slope for all the considered systems ($Q=0$, $Q=0.5$ and $Q=1$). This is expected, since this is a direct consequence of the two-body relaxation process. Still, the degree of mass segregation is different. The system with $Q=0$ emerges from the collapse with enhanced mass segregation. As explained, this is due to the mass segregation established in the substructures formed before the system collapses. This mass segregation is not completely removed after the clump merging process, and it is still evident in the dynamical state of the system just after the collapse.

A central massive object increases the velocity dispersion of the surrounding stars, reducing the formation of clumps and reducing the efficiency of mass segregation before and after the collapse. This explains why the curves of Fig. \ref{fig:Fig5} are closer to each ohter  than those in Fig. \ref{fig:Fig1}. The inclusion of a background gravitational potential, that mimics the presence of gas, smooths out two-body encounters, decreasing the efficiency of mass segregation and the formation of clumps, at least for $Q\geq 0.1$. The mass segregation that emerges in clumps before the collapse is also evident when studying systems with different number of stars (Fig. \ref{fig:Fig9}) and with different ratios between the number of massive to light stars (Fig. \ref{fig:Fig10}).

\section{Conclusions}
\label{sec:concl}

The main scope of this paper was the study of the, still open, topic of quick mass segregation in a stellar system composed by a moderate number of stars.
The relevance of this subject is suggested by observations showing, unexpectedly, mass segregation in young clusters, often still embedded in the gas left by the gaseous protocloud. Despite some effort, no firm conclusions on the origin of the observed mass segregation at the level observed in young embedded clusters have been drawn on a theoretical and/or numerical side. 

To attack this problem properly, and without exceeding in useless details which can mask the main physical processes governing the phenomenon, we considered clusters of moderate size ($128\leq N \leq 1,024$) as composed by stars with two different masses ($m_l$ and $m_h$, with $m_h=2m_l$), sampled from an initially homogeneous space distribution, spanning a range of initial virial ratios, $Q$, between cold conditions, $Q=0$, through cool, and warm conditions, $Q\leq 1$. For the case $N=1,024$ we varied the ratio of the number of heavy stars to that of light stars, $N_h/N_l$, in the range $0.125 \leq N_h/N_l \leq 1$, to have some representation of a realistic cluster mass function.

\begin{figure}
\centering
\includegraphics[width=\columnwidth]{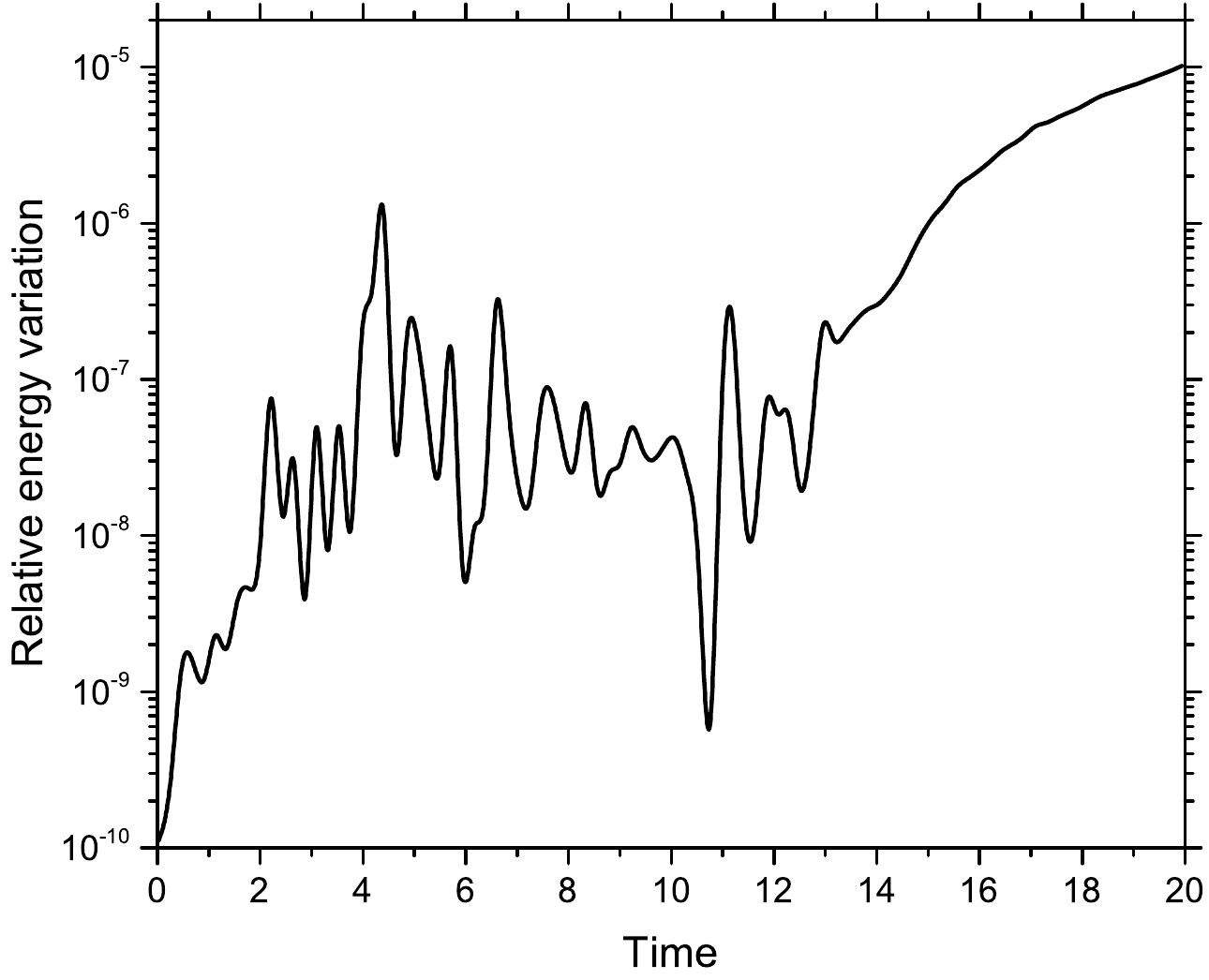}
\caption{ The  relative energy variation for the most critical case simulated: $N=1,024$, $Q=0$, the inclusion of a central stellar mass black hole and without a gaseous component} \label{fig:Fig13}
\end{figure}

Numerical simulations of the $N$-body evolutions were performed by mean of the high-precision and high-speed \texttt{HiGPUs} code \citet{dolcetta2013}.
It is worth noting that, although the code, at the moment, does not implement any special treatment to integrate neither binaries nor multiples and to deal accurately with very close encounters, the simulations we performed so far are numerically very precise despite the criticality of the initial conditions of the tested systems (initially null velocities of the stars, corresponding to the most violent collapse, $Q=0$). 
To quantify this, we show in Fig. \ref{fig:Fig13} the relative variation of the total energy of the system in 
the case $N=1,024$, $Q=0$, without presence of gas but with the presence of the central heavier particle. It is  
seen in this Figure that the relative variation of energy,
$|E(t)-E(0)|/|E(0)|$, is always  below $10^{-5}$ even after the core collapse of the system, that occurs at time $t_{cc}\sim 7t_c$, after which very tight multiple systems of stars form at the center, leading to a deep increase of the total energy of the system. This increase is mainly due to the significant reduction of the time step used in the integration of motion which yields to a deep increase of the number of iterations per time unit, and so a significant increase in the round-off error.
We have to note that the core collpase time and the characteristics of the emergin population of binaries would depend on the softening length assumed. Our assumption $\epsilon = 10^{-5}R$ means $\epsilon$ of the order of AU when $R=1$ pc, that means exclusions of tight binaries with a partial suppression of the quantity of kinetic energy released by binaries over the stellar environment.

We believe that the saturation of mass segregation at $T\simeq 10$ for runs with initial $Q=0$ is likely due to the formation of binaries, whose precise treatment requires particular care, out of the scopes of this paper.

We gave statistical relevance to our results by making the average of a set of sampled initial conditions for any of the input parameters chosen ($N$, $Q$, presence or absence of a black hole and of a background gaseous component).
Moreover, we quantified the level of mass segregation by the use of the powerful minimum spanning tree method  \citet{allison2009} coupled to the often used lagrangian radii. 
We have also taken into account the role of escapers on our results, accurately treating the escapers as those stars having positive individual mechanical energy and far enough from the cluster such to have almost zero probability to be recaptured by the cluster itself.

We found that clusters with cool initial conditions ($Q\lesssim 0.3$) segregate masses, in a way dependent on $N$ in the range of $N$ studied.
The dependence on $N$ is partly explained by the known relation between half-mass relaxation time and crossing time

 This happens in two distinct dynamical phases which both contribute to the mass segregation  on short or even very short time scales. They are: 
\begin{enumerate}
\item the formation of substructures (clumps) as the system undergoes its initial collapse, which segregate masses very quickly because of the small number of objects involved (this phenomenon acts in the first evolution of the system \emph{before} the bounce of the system);
\item the subsequent formation of a very dense core which is responsible for the mass segregation of the system that is clearly evident \emph{after} its bounce when no clumpy structures are formed.
\end{enumerate}
  
The first phase practically corresponds to the initial gravitational instability of the system and thus is related to the value of the Jeans mass of the system in dependence on the assumed initial virial ratio, in the sense that warmer initial conditions tend to cancel the role of this phase. On the other side, the second phase, although it is stronger in its effects for lower values of $Q$, remains significant in segregating masses on an extended interval of $Q$. This latter phase begins at about the moment of maximum compression of the system, whose following collisional evolution occurs on the crossing time of the quasi-virialized system.
In particular, for ~\lq cold\rq~ systems ($Q \lesssim 0.3$) the mass segregation observed  in the substructures, before the system collapses, is not completely erased after the collapse. Therefore, the dense subsystem that emerges after the collapse, shows a clear signature of mass segregation, despite it is not dynamically old enough to reach such degree of segregation yet.

On both the two phases of mass segregation has a relevance the inclusion of  a ~\lq background\rq~ potential which mimics the presence of a residual gas left after star formation as well as the inclusion of an additional body as heavy central object (a stellar black hole).
Quite clearly, the role on the cluster dynamics played by the background gas general potential depends on the characteristics of its time evolution (if it is very rapid it may influence both the phases).
The cases studied here (static and expanding gas) indicate that the addition of gravitational binding energy tends to smear out the local processes that give mass segregation, although segregation is still clearly visible.

The quick mass segregation for cool system seems to be a {\it robust} result in the sense it occurs in all the cases studied here, at different levels of efficiency in dependence on: i) the total number of stars in the cluster (the smaller the $N$ the more evident the mass segregation); ii) the number ratio of heavy to light stars in the cluster (the smaller this ratio the more evident the mass segregation).

The presence of a black hole ($m_{BH} = 25 m_h = 50 m_l$) 
influences mainly the second phase and the following, secular, evolution of the system. The black hole acts both in reducing the efficiency of close encounters between stars, thus decreasing the rate of energy exchanges and, so, the resulting mass segregation. 

To conclude: the results presented here are rough if considered on a pure astrophysical side, in what we did not pretend to represent in detail real open clusters, but  they are interesting enough on the side of dynamics of small- to intermediate- $N$-body systems.

Needed future steps to assess more firmly on the modes of violent mass segregation would require a proper inclusion of a primordial binary population, of a realistic mass function, a self consistent treatment of the residual gas and of the gas expelled via stellar winds during stars' evolution. Our future aims include also a very accurate treatment of close encounters between stars, employing a regularization method for the $N$-body problem. To do 
this, we will use an updated version of \texttt{HiGPUs}, named \texttt{HiGPUs-R}, which is still under development.
Finally, we think that the proper treatment of all these additional astrophysical {\it ingredients} will partially change the modes and properties of the violent mass segregation effect with respect to the present, simple, schematization, but at the same time we are reasonably convinced that the main features of this phenomenon as presented here will remain. 

\section{Acknowledgements}

We thank the referee, Mirek Giersz, for his comments and suggestions which greatly helped us to improve the paper.

\nocite{*}
\bibliography{spera_capuzzodolcetta}

\end{document}